\documentclass[11pt,twoside]{article} 
\usepackage{cspm-asp2006}
\usepackage{epsfig,graphicx,natbib,url}  
\usepackage{lscape} 
\begin{document}
\setcounter{page}{217}

\markboth{Tziotziou}{Cloud-model Inversion Techniques}
\title{Chromospheric Cloud-Model Inversion Techniques}
\author{Kostas Tziotziou}
\affil{National Observatory of Athens, Institute for Space Applications and Remote Sensing, Greece}

\begin{abstract}
    Spectral inversion techniques based on the cloud model are
    extremely useful for the study of properties and dynamics of
    various chromospheric cloud-like structures. Several
    inversion techniques are reviewed based on simple (constant source function)
    and more elaborated cloud models, as well as on grids of synthetic
    line profiles produced for a wide range of physical parameters by
    different NLTE codes. Several examples are shown of how such
    techniques can be used in different chromospheric lines, for the
    study of structures of the quiet chromosphere, such as
    mottles/spicules, as well as for active region structures such as
    fibrils, arch filament systems (AFS), filaments and
    flares.
\end{abstract}

\section{Introduction}

Observed intensity line profiles are a function of several
parameters describing the three-dimensional solar atmosphere, such
as chemical abundance, density, temperature, velocity, magnetic
field, microturbulence etc (which one would like to determine), as
well as of wavelength, space (solar coordinates) and time.
However, due to the large number of parameters that an observed
profile depends on, as well as data noise, model atmospheres have
to be assumed in order to restrict the number of these unknown
parameters. The term ``inversion techniques" refers to the
procedures used for inferring these model parameters from observed
profiles. We refer the reader to \citet{kt-mein00} for an extended
overview of inversion techniques. In this paper, we will review
only a class of such inversion techniques known in the solar
community as ``cloud models".

\begin{figure}
  \centering
  \includegraphics[width=0.4\textwidth]{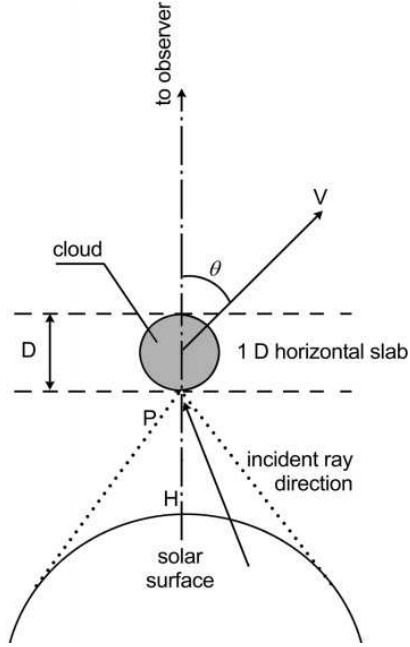}
  \caption[]{\label{kt-fig1}
  Geometry of the cloud model. $D$ is the geometrical thickness of
  the cloud at height $H$ above the solar surface and $V$ its
  velocity. From \citet{kt-hein}.
}\end{figure}

Cloud models refer to models describing the transfer of radiation
through structures located higher up from the solar photosphere,
which represents the solar surface, resembling clouds on earth's
sky (see Fig.~\ref{kt-fig1}). Such cloud-like structures, when
observed from above, would seem to mostly absorb the radiation
coming from below, an absorption which mostly depends on the
optical thickness of the cloud, that is the ``transparency" of the
cloud to the incident radiation and also on the physical
parameters that describe it. The possibility of observed emission
from such structures cannot, of course, be excluded when the
radiation produced by the cloud-like structure is higher than the
absorbed one. The aforementioned  processes are described by the
radiative transfer equation
\begin{equation}
I(\Delta\lambda) = I_{\rm 0}(\Delta \lambda)~
{\rm e}^{-\tau(\Delta\lambda)} + \int_{0}^{\tau(\Delta\lambda)} S_{t}~ {\rm e}^{-t(\Delta\lambda)}~ {\rm d}t\,, \label{kt-eq1}
\end{equation}
where $I(\Delta\lambda)$ is the observed intensity, $I_{\rm
0}(\Delta\lambda)$ is the reference profile emitted by the
background (the incident radiation to the cloud from below),
$\tau(\Delta\lambda)$ is the optical thickness and $S$ the source
function which is a function of optical depth along the cloud. The
first term of the right hand part of the equation represents the
absorption of the incident radiation by the cloud, while the
second term represents emission by the cloud itself.

The simple cloud model method introduced by \citet{kt-beck} arose
from the need to solve fast the radiation transfer equation and
deduce the physical parameters that describe the observed
structure. Beckers assumed that a) the structure is fully
separated from the underlying chromosphere, b) the source
function, radial velocity, Doppler width and the absorption
coefficient are constant along the line-of-sight (hereafter
LOS) and c) the background intensity is the same below the
structure and the surrounding atmosphere; hence it can be
extrapolated from a neighboring to the structure under study
region. Under the above assumptions the radiative transfer
equation is simplified to
\begin{equation}
I(\Delta\lambda) = I_{0}(\Delta \lambda)~
{\rm e}^{-\tau(\Delta\lambda)} + S (1 - {\rm e}^{-\tau(\Delta\lambda)})
\label{kt-eq2_1}
\end{equation}
and can be rewritten as
\begin{equation}
C(\Delta\lambda) = \displaystyle\frac{I(\Delta\lambda)- I_{0}(\Delta\lambda)}{I_{0} (\Delta\lambda)} \nonumber \\
 = \displaystyle\left(\frac{S}{I_{0} (\Delta\lambda)} - 1 \right) (1 -
{\rm e}^{-\tau(\Delta\lambda)})\,, \label{kt-eq2}
\end{equation}
where $C(\Delta\lambda)$ defines the contrast profile. A Gaussian
wavelength dependence is usually assumed for the optical depth as
follows
\begin{equation}
\tau(\Delta\lambda) = \tau_{0}~
{\rm e}^{-\displaystyle\left({\frac{\Delta\lambda-\Delta\lambda_{I}}
{\Delta\lambda_{\rm D}}}\right)^2}\,, \label{kt-eq3}  
\end{equation}
where $\tau_{0}$ is the line center optical thickness,
$\Delta\lambda_{I} = \lambda_{0} v/c$ is the Doppler shift
with $\lambda_{0}$ being the line center wavelength, $c$ the
speed of light and  $\Delta\lambda_{\rm D}$ is the Doppler width.
The latter depends on temperature $T$ and microturbulent velocity
$\xi_{\rm t}$ through the relationship
\begin{equation}
\Delta\lambda_{\rm D} = \frac{\lambda_{0}}{c}\sqrt{\xi^2_{\rm
t} + \frac{2 k T}{m}}\,,
\end{equation}
where $m$ is the atom rest mass. Other wavelength dependent
profiles than the Gaussian one can also be assumed for the optical
depth, e.g., a Voigt profile \citep{kt-tsir99}.

The four adjustable parameters of the model are the source
function $S$, the Doppler width $\Delta\lambda_{\rm D}$, the
optical thickness $\tau_{0}$ and the LOS velocity $v$. All
these parameters are assumed to be constant through the structure.
There are some crucial assumptions concerning Beckers' cloud model
(hereafter BCM):

\leftmargini=3ex
\leftmargini=3ex \begin{itemize} \itemsep=0ex \vspace{-1ex}

  \item the uniform background radiation assumption, which is not always true especially for cloud-like
  structures that do not reside above quiet Sun regions.
  Moreover, the background radiation plays an important role in the correct quantitative determination
  of the physical parameters.

  \item the neglect of incident radiation, the effects of which are of course not directly considered in BCM, but
  does play an important role in non-Local Thermodynamic Equilibrium (hereafter NLTE) modeling, since
  it determines the radiation field within the structure, that is the excitation and ionization conditions and
  hence the source function.

  \item the constant source function assumption which is not
  realistic especially in the optically thick case or not valid in
  the presence of large velocity gradients.

\end{itemize}
However, the cloud model works quite well for a large number of
optically thin structures and can provide useful, reasonable
estimates for the physical parameters that describe them. We refer
the reader to \citet{kt-alis90} for a detailed discussion on the
validity conditions of BCM for different types of contrast
profiles.

\section{Cloud Model Variants}

Since the introduction of the BCM method several improvements have
been suggested in the literature. When looking at the radiative
transfer equation (Eq.~\ref{kt-eq1}), it is obvious that all
efforts concentrate on a better description of the source function
$S$ which in BCM is considered to be constant. In the following
subsections some of these suggested improvements are described.

\subsection{Variable source function}
\label{kt-vsf}

\citet{kt-mein96} considered a source function that is a function
of optical depth and is approximated by a second-order polynomial
\begin{equation}
S_{\rm t} = S_0 + S_{\rm 1} \frac{\tau_0}{\tau_{\rm 0,
max}} + S_{2} \left( \frac{\tau_0}{\tau_{\rm 0,max}}
\right)^2  \label{kt-eq4}
\end{equation}
with the optical depth at the center of the line $\tau_0$
taking values between 0 and the total optical thickness at line
center $\tau_{\rm 0,max}$, while $S_0$, $S_{1}$ and
$S_{2}$ are functions of $\tau_{\rm 0,max}$. This formulation
was further improved by \citet{kt-hein}, who included also the
effect of cloud motion by assuming that $S_0$, $S_{1}$
and $S_{2}$ are now not only functions of $\tau_{\rm 0,max}$,
but also depend on the velocity $v$ of the structure.

\citet{kt-tsir99} assumed the parabolic formula
\begin{equation}
S_{t} = S_0 \left( 1 + \alpha \left( t - \frac{\tau_{\rm
0}}{2} \right)^2 \right) \label{kt-eq5}
\end{equation}
as an initial condition for the variation of the source function
with optical depth, where $S_0$ is the source function at
the middle of the structure, $\tau_0$ the optical depth at
line center, and $\alpha$ a constant expressing the variation of
the source function. However, their final results on the
dependence of the source on optical depth were in good agreement
with the results of \citet{kt-mein96}.

\begin{figure}
  \centering
  \includegraphics[width=0.42\textwidth]{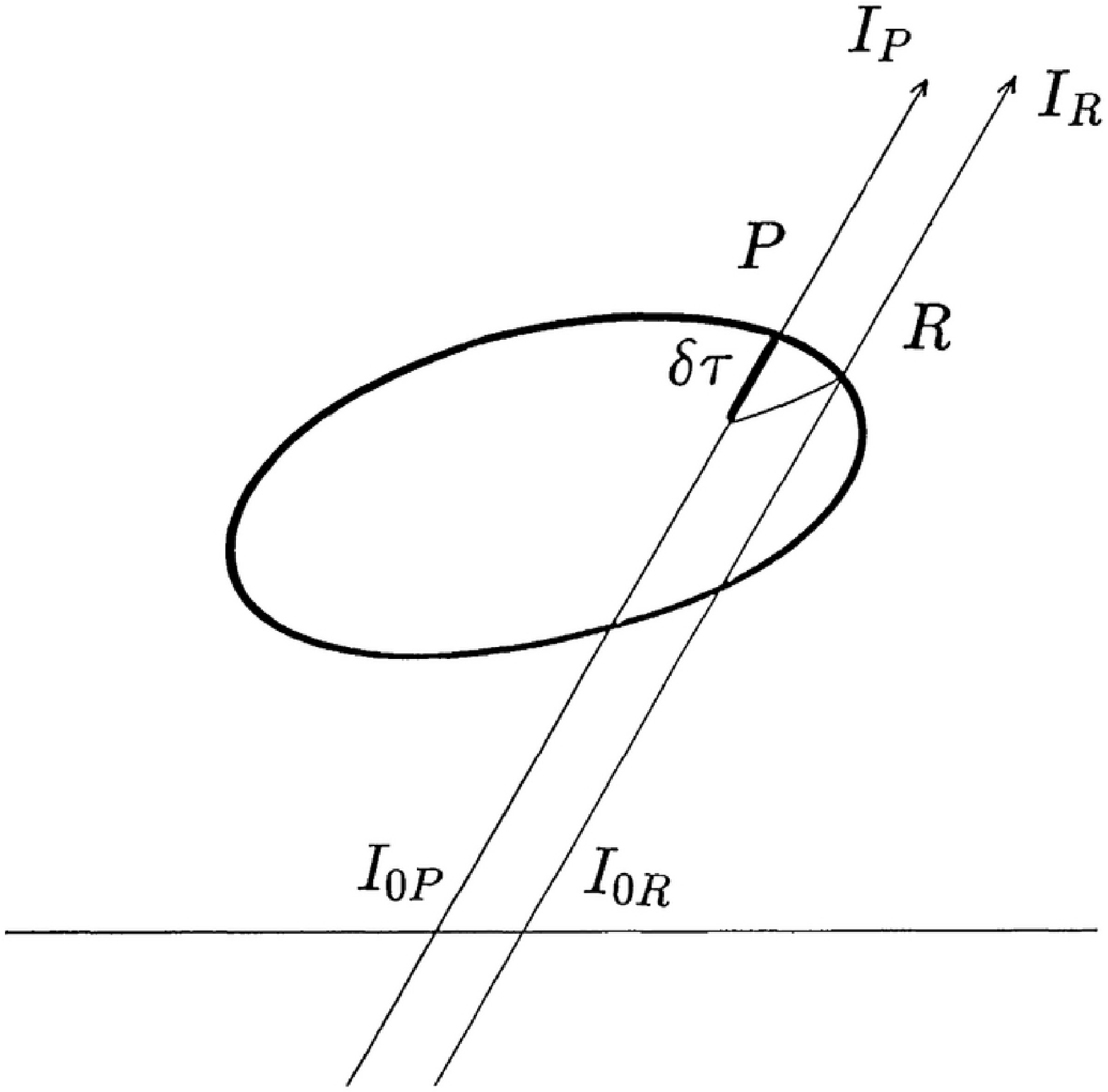}\hspace{0.5cm}
  \includegraphics[width=0.42\textwidth]{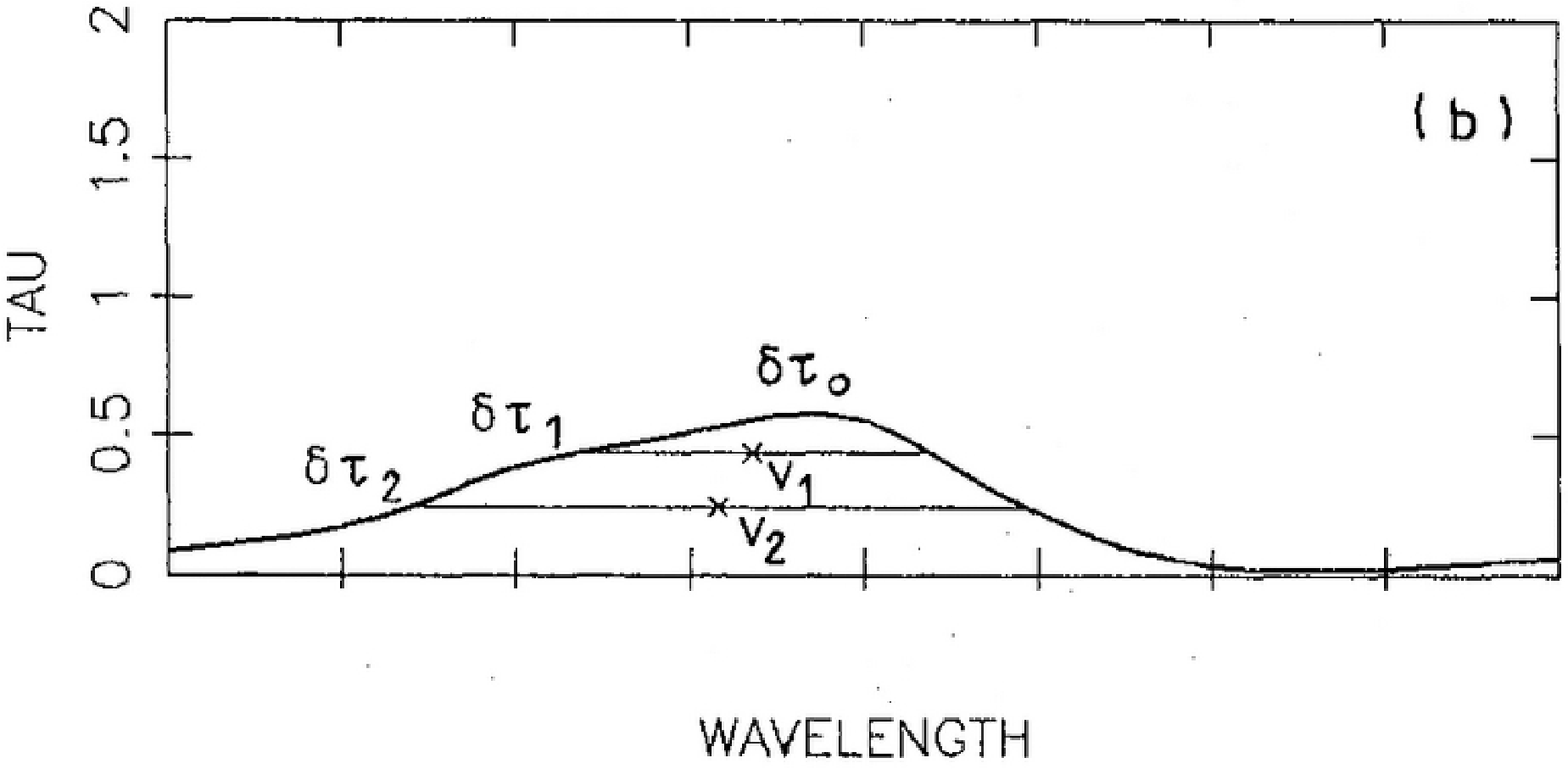}
  \caption[]{\label{kt-fig2}
  {\em Left:} Geometry of the cloud model in the case of first-order differential cloud model. From \citet{kt-hein92}.
  {\em Right:} The ``3-optical depths" procedure for solving the differential cloud model case which
  is described in Sect.~\ref{kt-solveDCM} (from \citet{kt-mein88}).
}\end{figure}

\subsection{Differential cloud models}
\label{kt-dcm}

First and second order differential cloud models (hereafter DCM1
and DCM2) were introduced by \citet{kt-mein88} to account for fast
mass flows observed on the disc, where BCM is not valid due to
fluctuations of the background and strong velocity gradients along the
LOS. DCM1 assumes that the source function $S$, temperature $T$ and
velocity $v$ are constant within a small volume contained between two
close lines of sight $P$ and $R$ (see Fig.~\ref{kt-fig2}, left
panel). If we assume that the variation of the background profile is
negligible ($I_{\rm 0P} \simeq I_{\rm 0R}$) for such close points then
the differential cloud contrast profile can be written as
\begin{equation}
C(P, R, \lambda) = \frac{I_{P}(\lambda)-I_{\rm
R}(\lambda)}{I_{R}(\lambda)} = \left(\frac{S}{I_{\rm
R}(\lambda)} - 1 \right)(1-{\rm e}^{-\delta\tau(\lambda)})
\label{kt-eq6}
\end{equation}
with
\begin{equation}
\delta\tau(\lambda) = \delta\tau_{0}~ {\rm e}^{- \displaystyle
\left( \frac{\lambda-\lambda_{0}-v\lambda_{\rm
0}/c}{\Delta\lambda_{\rm D}} \right)^2}\,,
\end{equation}
where $\Delta\lambda_{\rm D}$ is the Doppler width. The zero
velocity reference wavelength $\lambda_{0}$ is obtained by
averaging over the whole field of view. DCM1 is a method for
suppressing the use of the background radiation. If velocity
shears are present between neighboring LOS then DCM2 can be used
instead which requires the use of three neighboring LOS. We refer
the reader to \citet{kt-mein88} for the precise formulation of
DCM2 and to Table~1 of the same paper which summarizes the
validity conditions, constrains and results of the two models in
comparison to the classical BCM.

\subsection{Multi-cloud models}
\label{kt-mcmethod}

The multi-cloud model \citep{kt-gu92,kt-gu96} -- hereafter
MCM -- was introduced for the study of asymmetric, non-Gaussian
profiles, such as line profiles of post-flare loops, prominences
and surges and was based on the BCM and DCMs models. These
asymmetric line profiles are assumed to be the result of
overlapping of several symmetric Gaussian profiles along the LOS,
formed in small radiative elements (clouds) which have a)
different or identical physical properties and b) a source
function and velocity independent of depth. The profile asymmetry
mostly results from the relative Doppler shifts of the different
clouds. The total intensity $I_{\lambda}$ emitted by $m$
clouds is then given by the relation
\begin{equation}
I_{\lambda} = I_{0,\lambda} \, {\rm e}^{-\tau_{\lambda}} +
\sum_{j=1}^{m} S_{j} (1 - {\rm e}^{-\tau_{\lambda,j}}) \exp
\left(-\sum_{i=0}^{j-1} \tau_{\lambda,i} \right)\,,
\end{equation}
where $\tau_{\lambda,0} = 0$, $I_{0,\lambda}$ is the
background intensity, $\tau_{\lambda} =
\sum_{j=1}^{m}\tau_{\lambda,j}$ is the total optical depth of the
$m$ clouds and
\begin{equation}
\tau_{\lambda,j} =  \tau_{0,j}~ {\rm e}^{ - \displaystyle
\left(\frac{\lambda - \lambda_{0} -
\Delta\lambda_{0,j}}{\Delta\lambda_{{\rm D},j}}\right)^2 }
\end{equation}
$\Delta\lambda_{0,j} = \lambda_{0} v_{j}/c$, $S_{j}$,
$\tau_{0,j}$, $v_{j}$, $\Delta\lambda_{{\rm D},j}$ are
respectively the optical depth, Doppler shift, source function,
line-center thickness, velocity and Doppler width of the $j^{\rm
th}$ cloud.

A somewhat similar in philosophy, two-cloud model method was used
by \citet{kt-hein94} in their study of black and white mottles. It
was assumed that the LOS intersects two mottles treated as two
different clouds $c1$ and $c2$ with optical depths $\tau_{1}$ and
$\tau_{2}$ respectively. Hence the emerging intensity from the
lower mottle $I_{1}$ is assumed to be the background incident
intensity for the second upper mottle. Then, the equations
describing the radiation transfer through the two mottles are
\begin{eqnarray}
I_{2}(\Delta\lambda) & = & I_{\rm
1}~{\rm e}^{-\tau_{2}(\Delta\lambda)}
+ I_{c2}(\Delta\lambda) \nonumber\\
I_{1}(\Delta\lambda) & = & I_{\rm
0}~{\rm e}^{-\tau_{1}(\Delta\lambda)} + I_{c1}(\Delta\lambda)\,,
\end{eqnarray}
where $I_{0}$ is the background chromospheric intensity and
$I_{c1}$, $I_{c2}$ the intensity emitted by the two clouds
respectively. The novelty of the method is that for the
emitted by the clouds intensity, a grid of 140 NLTE models was
used which was computed for prominence-like structures by
\citet{kt-gout93}. So this method is a combination of MCM with
NLTE source function calculations which will be further discussed
in Section~\ref{kt-nlte}.

\subsection{The Doppler signal method}
\label{kt-dsmethod}

The Doppler signal method \citep{kt-geor90, kt-tsir00} can be used
when filtergrams at two wavelengths $-\Delta\lambda$ and
$+\Delta\lambda$ (blue and red side of the line) are available and
a fast determination of mass motions is needed. Then the Doppler
signal $DS$ can be defined from the BCM equations as
\begin{equation}
DS = \frac{\Delta I}{\sum I - 2 I_{0}} = \frac{{\rm e}^{-\tau^{+}} -
{\rm e}^{-\tau^{-}}}{2 - {\rm e}^{-\tau^{+}} - {\rm e}^{-\tau^{-}}}\,,
\end{equation}
where $\Delta I = I(-\Delta\lambda) - I(+\Delta\lambda)$, $\sum I
= I(-\Delta\lambda) + I(+\Delta\lambda)$ and $\tau^{\pm} =
\tau(\pm\Delta\lambda)$. The Doppler signal $DS$ has the same sign
as velocity and can be used for a qualitative description of the
velocity field. The left hand side of the above equation can be
determined by the observations while the right hand clearly does
not depend on the source function. Quantitative values for the
velocity can be obtained when $\tau_{0} < 1$; then the Doppler
signal equation reduces to
\begin{equation}
DS =  \frac{\tau^{-} - \tau^{+}}{\tau^{-} + \tau^{+}}
\end{equation}
and the velocity $v$ -- once $DS$ is calculated from the
observations and a value of the Doppler width $\Delta\lambda_{\rm
D}$ is obtained from the literature or assumed -- is given by the
equation
\begin{equation}
v = \displaystyle \frac{\Delta\lambda_{\rm D}^2}{4\Delta\lambda}
\frac{c}{\lambda} \ln \left( \frac{1 + DS}{1 - DS} \right)\,.
\end{equation}

\subsection{Avoiding the background profile}
\label{kt-nobg}

\citet{kt-liu01} in order to avoid the use of the background
profile needed in BCM assumed that it is symmetric, that is
$I_{0}(\Delta\lambda) = I_{0}(-\Delta\lambda)$. Then it
can easily be shown that we can obtain the relationship
\begin{equation}
\Delta I(\Delta\lambda) = I(\Delta\lambda) - I(-\Delta\lambda) =
[I(\Delta\lambda)-
S][1-{\rm e}^{\tau(\Delta\lambda)-\tau(-\Delta\lambda)}]\,,
\end{equation}
which does not require the use of the background for the
derivation of the physical parameters.

\subsection{NLTE methods}
\label{kt-nlte}

\begin{figure}
  \centering
  \includegraphics[width=0.52\textwidth]{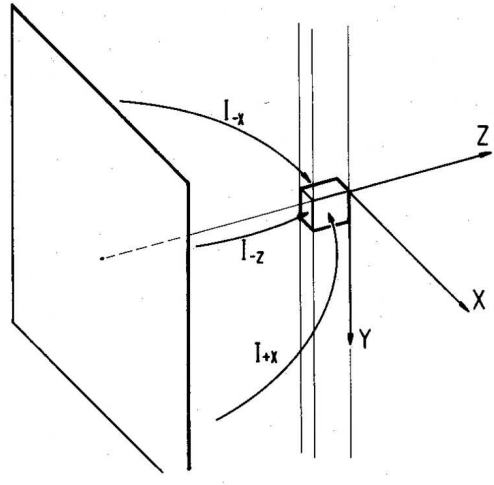}
  \caption[]{\label{kt-fig13}
  Geometry of a two-dimensional cloud model slab. The incident radiation comes not only from below, but
  also from the sides of the structure. From \citet{kt-vial82}.
}\end{figure}

As Eq.~\ref{kt-eq1} shows, in the general case, the source
function $S$ within a cloud-like structure is not constant, but
usually depends on optical depth. In order to calculate this
dependence, the NLTE radiative transfer problem within the
structure has to be solved, taking into account all excitation and
ionization conditions within the structure. Several efforts have
been undertaken in the past for such NLTE calculations, usually
for the case of filaments or prominences. Such NLTE calculations
started from the one-dimensional regime, where the cloud-like
structure is approximated by an infinite one-dimensional slab (see
Fig.~\ref{kt-fig1}) or a cylinder. We refer the reader to the
works of \citet{kt-heas74}, \citet{kt-heas76}, \citet{kt-heas76b},
\citet{kt-mozo78}, \citet{kt-font85}, \citet{kt-hein87},
\citet{kt-gout93}, \citet{kt-hein95}, \citet{kt-gout04} for an
overview of such one-dimensional NLTE models. The philosophy of
two-dimensional NLTE models is similar to the one-dimensional
models, but now the cloud-like structure is replaced by a
two-dimensional slab or cylinder which is infinite in the third
dimension, allowing both vertical, as well as horizontal radiation
transport (see Fig.~\ref{kt-fig13}). Furthermore, the incident
radiation is treated as anisotropic and comes now not only from
below, but also from the sides of the structure. We refer the
reader to the works of \citet{kt-miha78}, \citet{kt-vial82},
\citet{kt-pale93}, \citet{kt-auer94}, \citet{kt-hein01},
\citet{kt-gout05} for an overview of such two-dimensional NLTE
models.

A general recipe for such NLTE models, which is modified according
to the specific needs, i.e.\ the line profile used and the
structure observed, has as follows:
\leftmargini=3ex \begin{itemize} \itemsep=0ex \vspace{-1ex}

  \item The cloud-like structure is assumed to be a 1-D or 2-D slab or
  cylinder at a height $H$ above the photosphere. This slab/cylinder
  can be considered to be either isothermal \citep[e.g.,][]{kt-hein95}
  or isothermal and isobaric \citep[e.g.,][]{kt-pale93}.

  \item The incident radiation comes in the case of 1-D models only
   from below and in the case of 2-D models also from the sides and
   determines the radiation field within the structure, that is all
   excitation and ionization conditions.

  \item A multi-level atom plus continuum is assumed. The larger the
  number of atomic levels used, the more computationally demanding the
  method is. Complete or partial redistribution effects (CRD or PRD)
  are also assumed depending on the formation properties of the
  line. Methods with CRD are computationally much faster so sometimes
  CRD is used but with simulated PRD effects taken into account
  \citep[e.g.,][]{kt-hein95}.

  \item Some physical parameters are assigned to the slab/cylinder,
  like temperature $T$, bulk velocity $v$, geometrical thickness $Z$,
  electronic density $N_{\rm e}$ or pressure $p$. Calculations with
  electronic density are usually faster than calculations with
  pressure.

  \item The radiative transfer statistical equilibrium equations are
  numerically solved and the population levels are found and hence the
  source function as a function of optical depth for a set of selected
  physical parameters.

\end{itemize}
Once the source function $S$ is obtained as a function of optical
depth, Eq.~\ref{kt-eq1} can be solved in order to calculate the
emerging observed profile from the structure which is going to be
compared to the observed one.

\section{Solving the Cloud Model Equations}

In the following subsections some of the methods used to solve the
cloud model equations are reviewed. We remind the reader that
whenever the background profile is needed, either the average
profile of a quiet Sun region is taken or the average profile of a
region close to the structure under study.

\subsection{Solving the constant-$S$ case with the ``5-point'' method}
\label{kt-5point}

\citet{kt-mein96} introduced the ``5-point method" for solving the
BCM equation with constant $S$. According to this method five
intensities of the observed and the background profile at
wavelengths $\lambda_{1}$, $\lambda_{2}$ (blue wing of the
observed profile), $\lambda_{3}$, $\lambda_{4}$ (red wing
of the the observed profile) and the line-center wavelength
$\lambda_{0}$ are used for solving Eqs.~\ref{kt-eq2} and
\ref{kt-eq3}. It is an iterative method that works as follows:
\leftmargini=3ex
\leftmargini=3ex \begin{itemize} \itemsep=0ex \vspace{-1ex}

  \item The line-center wavelength $\lambda_{0}$ profile and background intensities
  are used for calculating $S$, where $\tau_{0}$, $\Delta\lambda_{\rm D}$ and $v$ are determined in
  a previous iteration. At the first step of the iteration some values can be
  assumed and $S$ can be taken as equal to zero.

  \item Profile and background intensities at wavelengths $\lambda_{\rm
  1}$ and $\lambda_{3}$ are used for calculating a new $\tau_{\rm
  0}$.

  \item Afterwards a new $\Delta\lambda_{\rm D}$ is calculated using the
  other two remaining wavelengths $\lambda_{2}$ and $\lambda_{4}$.

  \item Finally a new velocity is calculated from wavelengths $\lambda_{1}$,
  $\lambda_{2}$, $\lambda_{3}$, and $\lambda_{4}$ and then
  a reconstructed profile obtained using the derived parameters which is
  compared to the observed one. If any of the departures between the reconstructed
  and the observed profile is higher than an assumed small threshold value
  (i.e.\ 10$^{-4}$) then the aforementioned procedure is repeated until
  convergence is achieved. If no convergence is gained after a certain number
  of iterations then it is assumed that no solution exists.

\end{itemize}
We refer the reader to \citet{kt-mein96} for a detailed
description of the analytical equations described above.

\subsection{Solving the constant-$S$ case with an iterative least-square fit}

This method which was used by \citet{kt-alis90} and further
described in \citet{kt-tsir99} and \citet{kt-tzio03} fits the
observed contrast profile with a curve that results from an
iterative least-square procedure for non-linear functions which is
repeated until the departures between computed and observed
profiles are minimized. The coefficients of the fitted curve are
functions of the free parameters of the cloud model. At the
beginning of the iteration procedure initial values have to be
assumed for the free parameters and especially for the source
function $S$ which is usually estimated from some empirical
approximate expressions that relate it to the line-center
contrast. This method is very accurate and usually converges
within a few iterations. The more observed wavelengths used within
the profile, the better the determination of the ohysical
parameters is. However, as \citet{kt-tzio03} have reported, the
velocity calculation can overshoot producing very high values, if
the wings of the profile are not sufficiently covered by observed
wavelengths. The suggested way to overcome the problem is to
artificially add two extra contrast points near the continuum of
the observed profile where the contrast should be in theory equal
to zero.

This iterative method can also be successfully used not only in
the case of a constant source function $S$, but also for cases
with a prescribed expression for the source function, such as the
parabolic expression of Eq.~\ref{kt-eq5} used by
\citet{kt-tsir99}.

\subsection{Solving the constant-$S$ case with a constrained nonlinear
least-square fitting technique}

The constrained nonlinear least-square fitting technique, used by
\citet{kt-chae06} for the inversion of a filament with BCM, was
introduced by \citet{kt-chae98}. According to the method a)
expectation values $p_{i}^{e}$ of the $i$th free parameters, b) their
uncertainties $\varepsilon_{i}$, as well as c) the data to fit are
provided ($M$ wavelengths along the profile) and then a set of $N$
free parameters ${\bf p}=(p_{0}, p_{1}, ...p_{N-1})$ are
sought, i.e.\ ${\bf p} = (S, \tau_{0}, \lambda_{0},
\Delta\lambda_{\rm D})$, that minimize the function
\begin{equation}
H({\bf p}) = \sum_{j=0}^{M-1} \left( \frac{C^{\rm obs}_{j} -
C^{\rm mod}_{j}({\bf p})}{\sigma_{j}} \right)^2 + \sum_{i=0}^{N-1}
\left( \frac{p_{i} - p^{e}_{i}}{\varepsilon_{i}} \right)^2\,,
\end{equation}
where $C^{\rm obs}_{j}$ and $C^{\rm mod}_{j}$ are respectively the
observed and calculated with the expectation values contrasts and
$\sigma_{j}$ the noise in the data. The first term of the sum $H$
represents the data $\chi^2$, while the second term the
expectation $\chi^2$ which regularizes the solution by
constraining the probable range of free parameters. For very small
values of $\varepsilon_{i}$ the solution will not be much
constrained by the data and will be close to the chosen set of
expectation values $p_{i}^{e}$, while for large values of
$\varepsilon_{i}$ it will be mostly constrained by the data and
not by the expectation values. We refer the reader to
\citet{kt-chae06} for a detailed discussion of the effects of
constrained fitting.

\subsection{Solving the variable-$S$ case}
Apart from the iterative least-square procedure described above
which can be used when the source function varies in a prescribed
way, \citet{kt-mein96} have introduced also the ``4-point method"
for solving the case of a source function that is described by the
second order polynomial of Eq.~\ref{kt-eq4}. According to the
method an intensity $I'(\Delta\lambda)$ can be defined as follows
\begin{eqnarray}
I'(\Delta\lambda) & = & I(\Delta\lambda) -
\frac{1-[\tau(\Delta\lambda)+1]
~{\rm e}^{-\tau(\Delta\lambda)}}{\tau(\Delta\lambda)}~ S_{1} +
\nonumber
\\ &  & \frac{2-[\tau^2(\Delta\lambda) + 2 \tau(\Delta\lambda) +
2]~{\rm e}^{-\tau(\Delta\lambda)}} {\tau^2(\Delta\lambda)}~ S_{2}
\end{eqnarray}
and then the radiative transfer equation reduces to
\begin{equation}
I'(\Delta\lambda) = S_0 + (I_{0} - S_{\rm
0})~{\rm e}^{-\tau(\Delta\lambda)}\,.
\end{equation}
This equation can be solved now using the iteration procedure
described in Sect.~\ref{kt-5point}, with the modification that
$I(\Delta\lambda)$ is now replaced by $I'(\Delta\lambda)$ and that
the source function calculation in the first step is replaced by
the assumed theoretical relation for $S$ given by
Eq.~\ref{kt-eq4}.

\subsection{Solving the DCM cases}
\label{kt-solveDCM}

A method for solving the differential cloud model cases is the
``3-optical depths" procedure introduced by \citet{kt-mein88}.
According to this procedure: \leftmargini=3ex
\leftmargini=3ex \begin{itemize} \itemsep=0ex \vspace{-1ex}
  \item the zero velocity reference is obtained from
  the average profile over the whole field of view;
  \item a value $S$ is assumed between zero and the line-center
  intensity (in principle it could even work also for
  emitting clouds) and a function $\delta\tau(\lambda)$ is
  derived from Eq.~\ref{kt-eq6}. The latter is characterized by the
  maximum value $\delta\tau_0$ and $\delta\tau_{1}$,
  and the values $\delta\tau_{2}$ (see right panel of Fig.~\ref{kt-fig2}) which correspond to
  the half widths $\Delta\lambda_{1}$ and $\Delta\lambda_{2}$
  respectively and are given by the following relations
  \begin{eqnarray}
        \delta\tau_{1} & = & \delta\tau_{\rm
        0}~{\rm e}^{-(\Delta\lambda_{1}/\Delta\lambda_{\rm D})^2}
        \nonumber \\
        \delta\tau_{2} & = & \delta\tau_{\rm
        0}~{\rm e}^{-(\Delta\lambda_{2}/\Delta\lambda_{\rm D})^2}\,;
        \label{kt-eq17}
  \end{eqnarray}
  \item the code fits $S$ and $\Delta\lambda_{\rm D}$ by the conditions of
  Eq.~\ref{kt-eq17} coupled with Eq.~\ref{kt-eq6} and the solutions are
  assumed to be acceptable when the radial velocities $v_{1}$
  and $v_{2}$, which correspond to widths $\Delta\lambda_{1}$ and $\Delta\lambda_{2}$
  respectively and are defined as the displacement of the middle
  of these chords compared with the zero reference position,
  are not that different. When convergence is achieved the $\delta\tau(\lambda)$
  curve is well represented by a Gaussian and the Doppler width
  $\Delta\lambda_{\rm D}$ is independent of the chord
  $\Delta\lambda$.
\end{itemize}

\subsection{Solving the MCM case}

We refer the reader to the papers by \citet{kt-li92} and
\citet{kt-li93,kt-li94} for a detailed description of the methods
and mathematical manipulations used for fitting observed profiles
with the multi-cloud method, which unfortunately are not easy to
concisely describe within a few lines.

\subsection{Using NLTE Methods}

The most straightforward method for deriving the parameters of an
observed structure with NLTE calculations would be the calculation
of a grid of models for a wide range of the physical parameters
used to describe the structure. However, the calculation of such a
grid is computationally demanding, especially in the case when a)
a large number of atomic levels is assumed  and/or b) partial
redistribution effects (PRD) are taken into account and/or c) a
two-dimensional geometry is considered. For such cases, either a
very small grid of models is constructed and thus only approximate
values for the observed structure are derived or ``test and try"
methods are used where the user makes a ``good guess" for the
physical parameter values, proceeds to the respective NLTE
calculations, compares the derived profile(s) with the observed
one(s) and applies the necessary adjustments to the model
parameters according to the derived results.

However, nowadays the construction of a large grid of models,
although time-demanding, becomes more of a common practice with
the extended capabilities of modern computers. We refer the reader
to \citet{kt-molo99} and \citet{kt-tzio01} for two such examples,
both considering a one-dimensional isothermal slab for a
cloud-like structure, which is the same filament observed and
studied in the H$\alpha$ in \ion{Ca}{ii} 8542\,\AA\, lines respectively. The general
methodology used in the case of grids of models is the following:
\leftmargini=3ex \begin{itemize} \itemsep=0ex \vspace{-1ex}
  \item a grid of synthetic line profiles for a wide range of
  model parameters is computed using NLTE calculations for the source function, as
  described in Sect.~\ref{kt-nlte};
  \item these synthetic profiles are convolved with the characteristics
  of the instrument used for the observations in order to simulate
  its effects on the observed profiles;
  \item each observed profile is compared with the whole
  library of convolved synthetic profiles and the best
  fit is derived, that is the synthetic profile with the smallest departure, and hence the
  physical parameters that describe it;
  \item an interpolation (linear or parabolic) between neighboring points in
  the parameter space can also be used, for a more accurate quantitative determination of the
  physical parameters that best describe the observed profile.
  \end{itemize}

Grid models based on NLTE calculations have many advantages since
preferred geometries, temperature structures, etc can be used, no
iterations are required, errors can be easily defined from the
parameter space and inversions are nowadays becoming faster with
modern computers.

\section{Validity of the Cloud Models}

The validity of the cloud model used for an inversion obviously
strongly depends on a) the method used, b) the assumptions that
were made for the model atmosphere describing the structure and c)
the specific characteristics of the structure under study. Most of
the reviewed papers in Sect.~\ref{kt-examples}, concerning
applications of different cloud models, have extended discussions
on the validity of the cloud model method and the results
obtained, as well as the limitations of the method for the
specific structure. However, below, some studies found in
literature about the validity of cloud models are presented.

\begin{figure}
  \centering
  \includegraphics[width=0.5\textwidth]{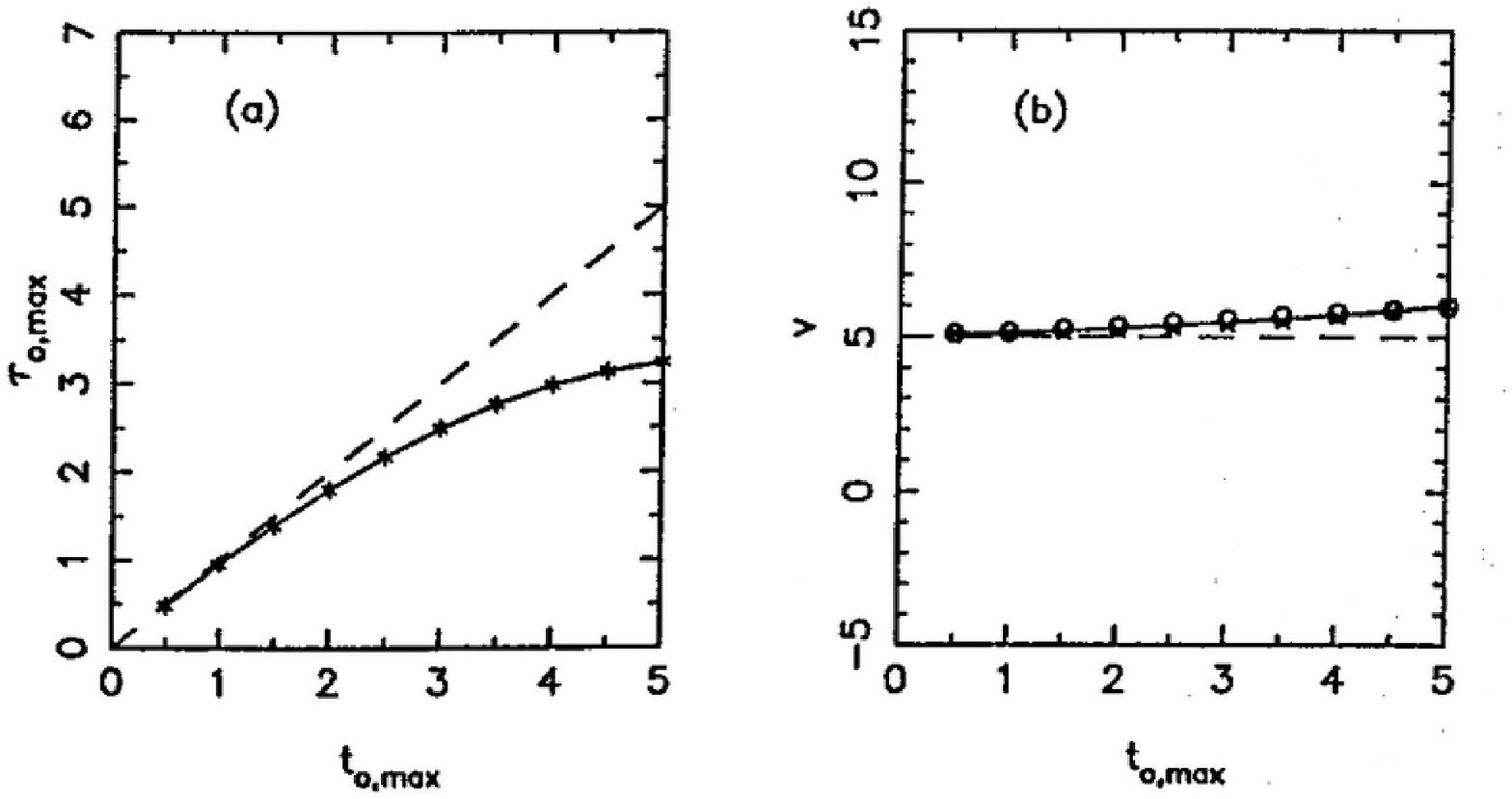}
  \includegraphics[width=0.48\textwidth]{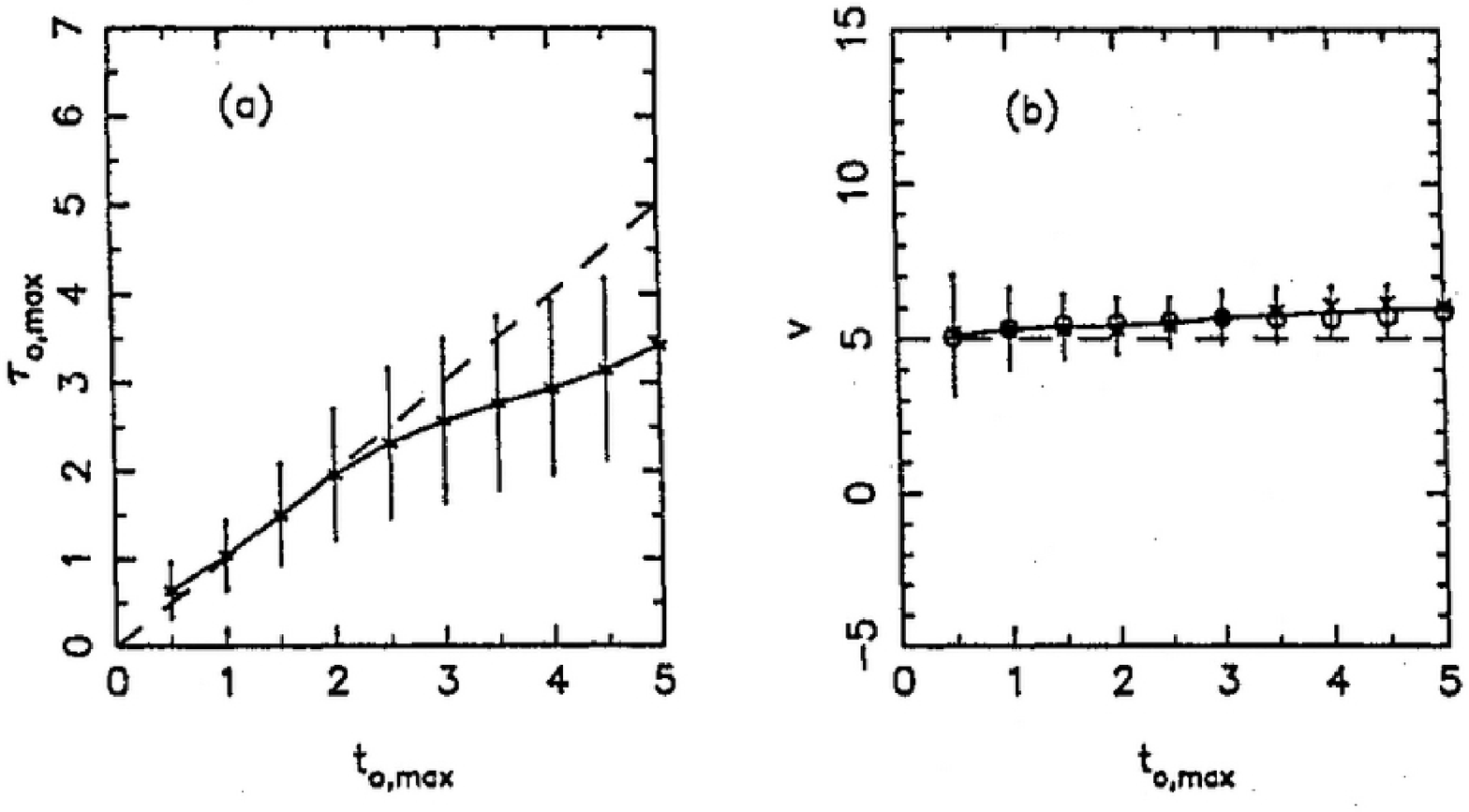}
  \caption[]{\label{kt-fig3}
  {\em Two left panels:} The calculated optical depth $\tau_{\rm 0,max}$
  and velocity $v$ with BCM (constant source function) versus the
  assumed optical thickness. The dashed curve is the model, the solid curve
  the inversion. {\em Two right panels:} Same plots but with
  added Gaussian noise. From \citet{kt-mein96}.
}\end{figure}

\begin{figure}
  \centering
  \includegraphics[width=0.49\textwidth]{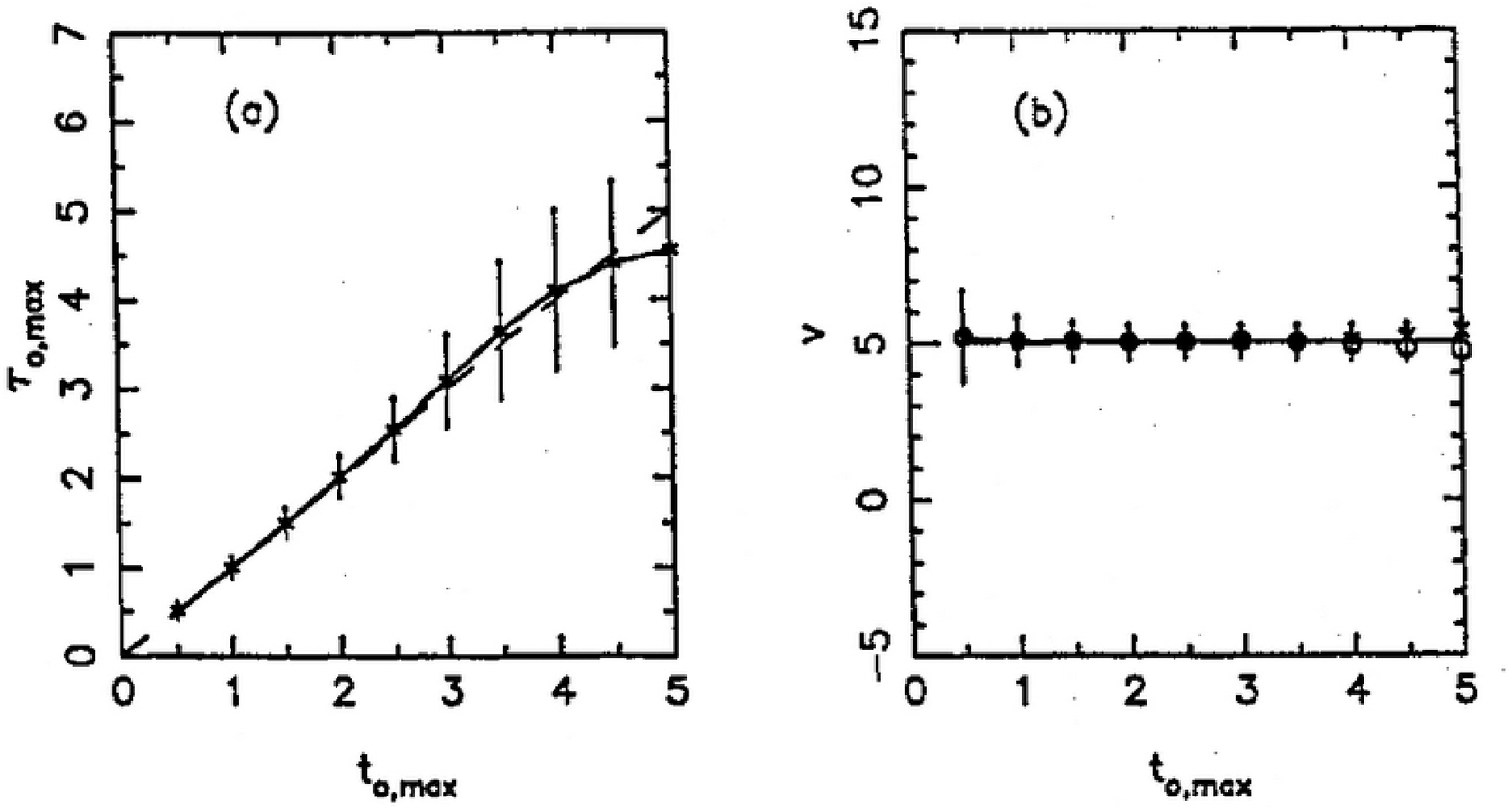}
  \includegraphics[width=0.49\textwidth]{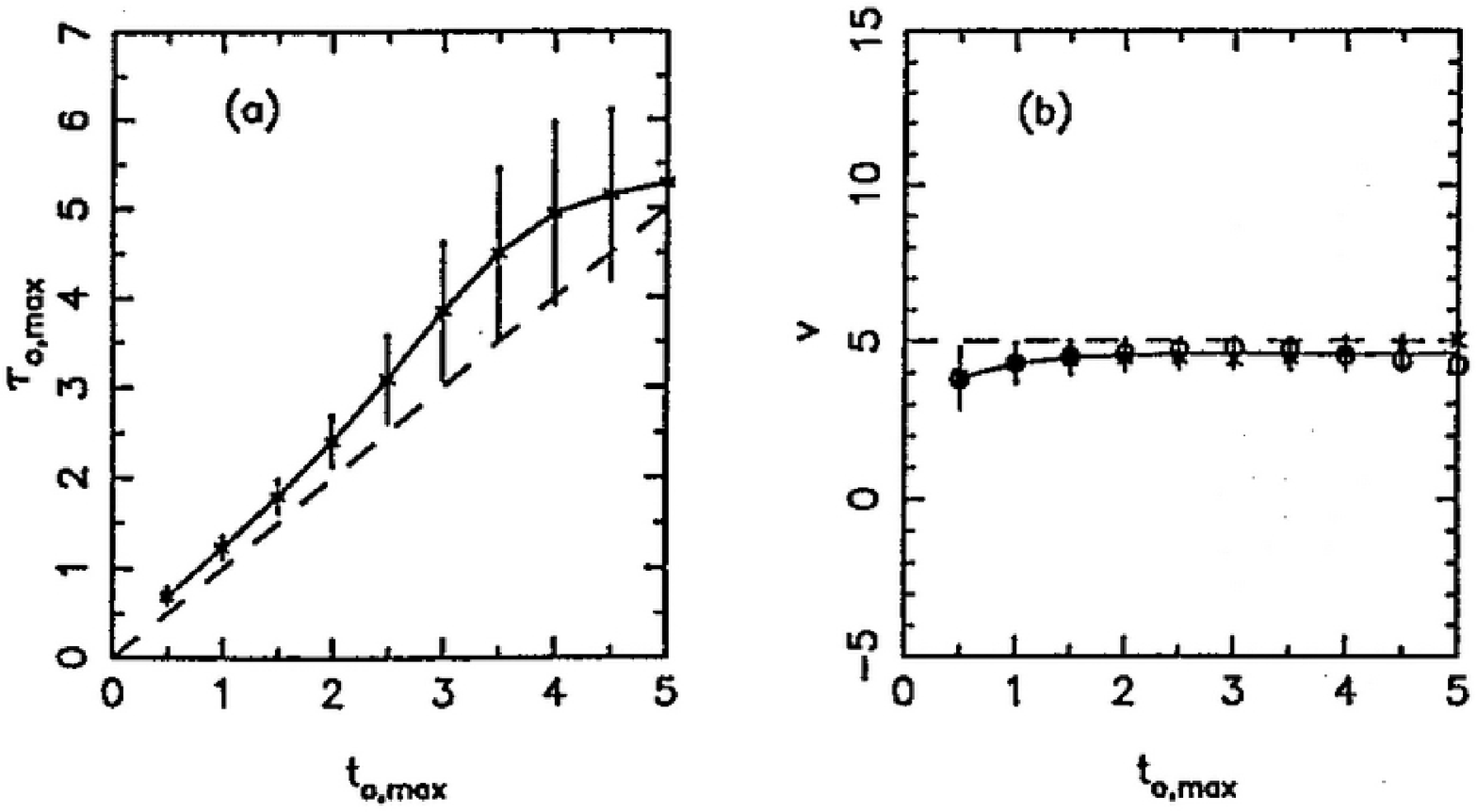}
  \caption[]{\label{kt-fig4}
  {\em Two left panels:} The calculated optical depth $\tau_{\rm 0,max}$ and velocity $v$ using a cloud model
  with variable source function (see Eq.~\ref{kt-eq4}) depending only on line-center
  optical thickness versus the assumed optical thickness.  Gaussian noise
  has also been taken into account. {\em Two right panels:} Same plots
  but for an over-estimated chromospheric background profile. From \citet{kt-mein96}.
}\end{figure}

\citet{kt-mein96} presented a rather detailed study about the
validity of BCM (constant source function), as well as of cloud
models with a variable source function as described in
Eq.~\ref{kt-eq4} (depending only on line-center optical thickness)
by inverting theoretical profiles produced with a NLTE code and
comparing the resulting model parameters from the inversion with
the assumed ones. Figure~\ref{kt-fig3} ({\rm two left panels})
shows the results of the inversion versus the assumed model
optical thickness for the BCM inversion (constant source
function). The calculated optical thickness is smaller, with the
difference increasing with the thickness of the cloud, while the
difference in velocity is no more than 20\% and only for high
values of the thickness. The figure shows that for optically thin
structures there is practically no difference in the obtained
results. When noise is included (Fig.~\ref{kt-fig3}, {\rm two
right panels}) the error increases for increasing thickness but
the mean values stay almost the same. Again for optically thin
structures the difference in the results is very small.

\begin{figure}
  \centering
  \includegraphics[width=0.75\textwidth]{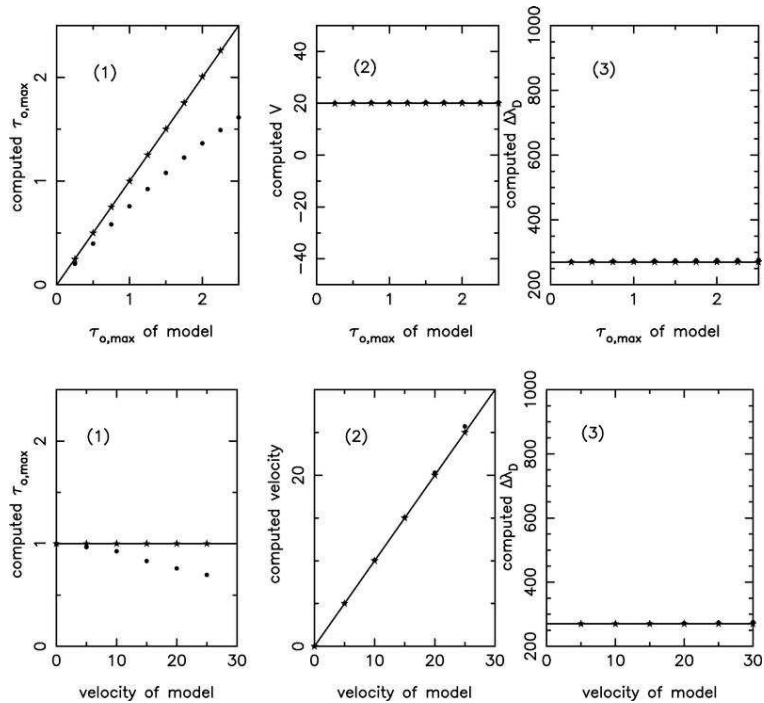}
  \caption[]{\label{kt-fig5}
  Comparison of the results obtained with method (a) represented by dots and with method (b)
  represented by asterisks (see text for details of the methods) with the assumed model values
  (solid curve). From \citet{kt-hein}.
}\end{figure}

Figure~\ref{kt-fig4} ({\rm two left panels}) shows the results of
the inversion versus the assumed model optical thickness for a
cloud model with variable source function according to
Eq.~\ref{kt-eq4} depending only on line-center optical thickness
with an added Gaussian noise; without noise the results are
perfectly reproduced. We see that the differences are now almost
negligible for a large range of the assumed optical thickness and
the parameters are better determined. However, when taking a
slightly brighter background (Fig.~\ref{kt-fig4}, two right
panels) we see that the calculated values of the optical thickness
are larger than the assumed ones, while the estimation of velocity
is still rather good. This shows the importance of a correct
background profile choice in cloud model calculations.

\begin{figure}
  \centering
  \includegraphics[width=0.32\textwidth]{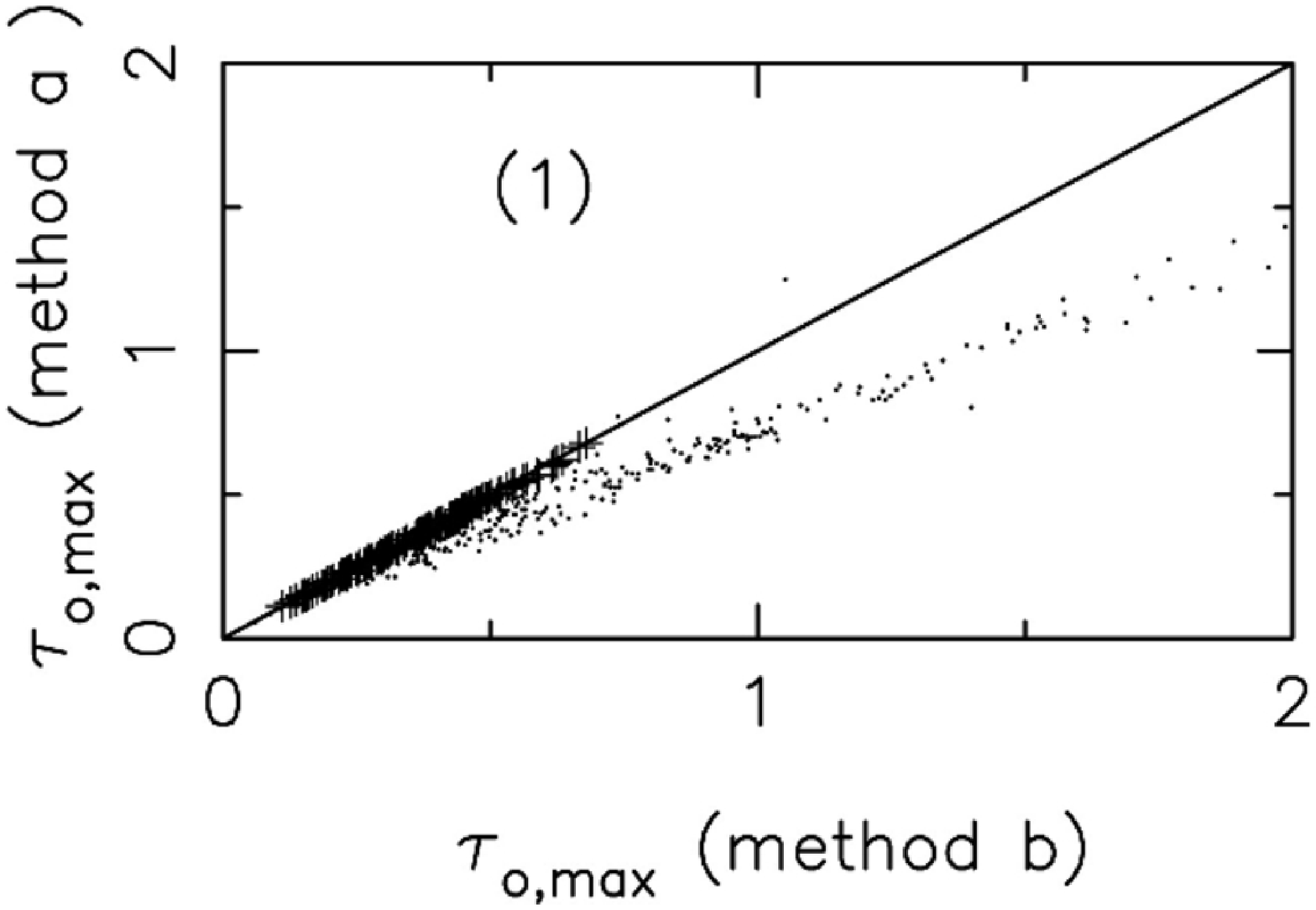}
  \includegraphics[width=0.32\textwidth]{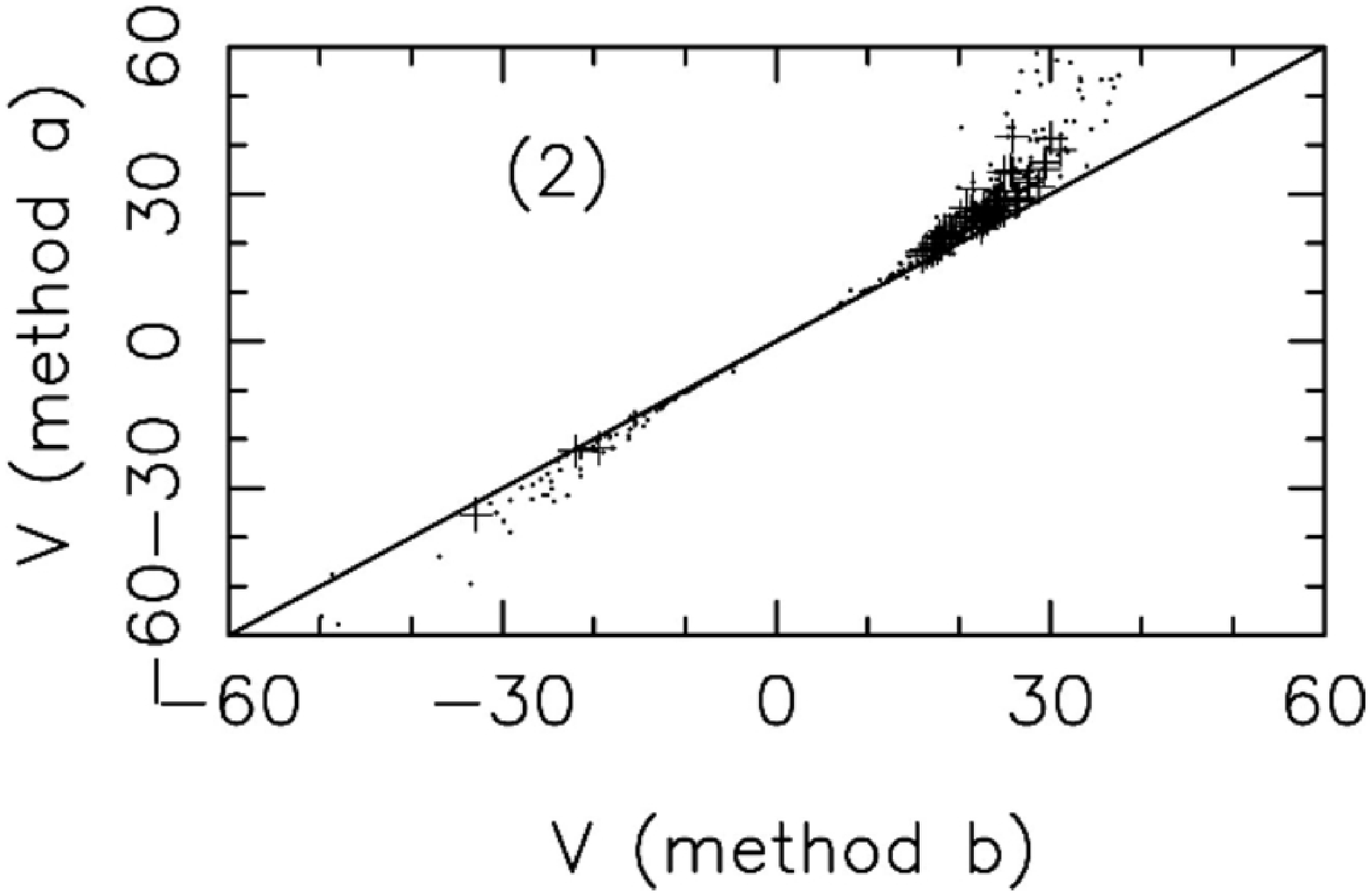}
  \includegraphics[width=0.32\textwidth]{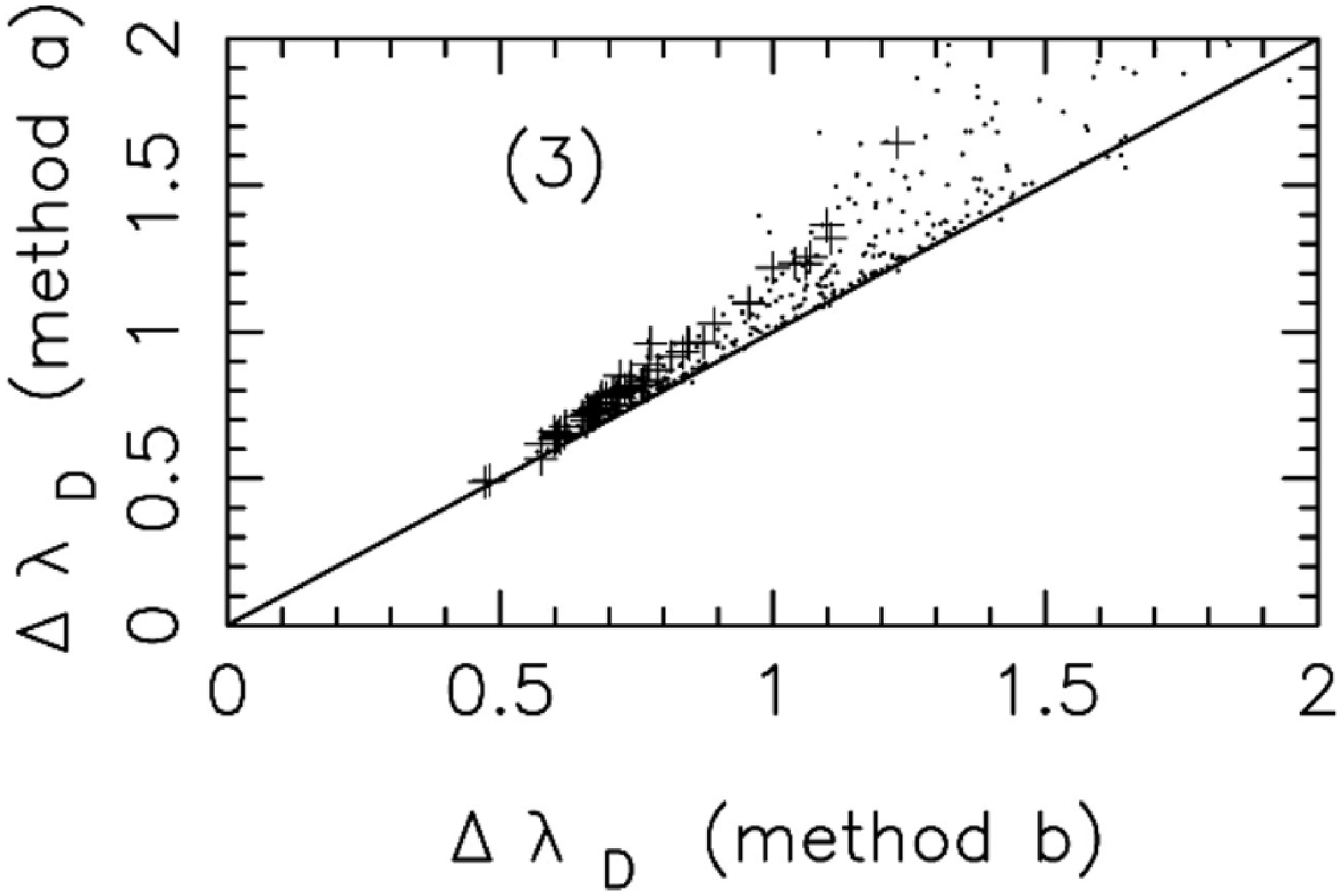}
  \caption[]{\label{kt-fig6}
  Comparison of the results obtained with the two methods (a) and (b) (see text for details)
  from the inversion of observed profiles of a dark arch filament. Scatter plots are shown for (1) optical
  thickness, (2) velocity (in~km~s$^{-1}$), and (3) Doppler width (in\,\AA).
  From \citet{kt-hein}.
}\end{figure}

\citet{kt-hein} has repeated the same exercise (inversion of NLTE
synthetic profiles) for a cloud model with a variable source
function according to Eq.~\ref{kt-eq4} depending a) only on
line-center optical thickness (method a) and b) on line-center
optical thickness and velocity (method b). Some of their results
are shown in Fig.~\ref{kt-fig5}. We see that although with method
(a) there are some differences in the calculation of optical
thickness, similarly to \citet{kt-mein96}, method (b) gives exact
solutions. \citet{kt-hein} have also applied the two methods in
observed profiles of a dark arch filament. Figure~\ref{kt-fig6}
shows the comparison of the results obtained with the two methods.

We refer also the reader selectively to a) \citet{kt-molo99}
(Fig.~12 of their paper) for a comparison of inversion results for
a filament with a NLTE method and a cloud model with a parabolic
$S$, b) \citet{kt-schm03} (Fig.~16 of their paper) for a
comparison of inversion results for a filament with a NLTE method
and a constant source function cloud model, c) \citet{kt-tsir99}
(Fig.~5 of their paper) for a comparison of inversion results for
mottles for cloud models with a constant and parabolic $S$, and d)
\citet{kt-alis90} (Fig.~8 to 11 of their paper) for a comparison
of inversion results for an arch filament system  with Beckers'
cloud model, the Doppler signal method and the differential cloud
model.

\section{Examples of Cloud Model Inversions}
\label{kt-examples}

Cloud models have been so far successfully applied for the
derivation of the parameters of several cloud-like solar
structures of the quiet Sun, such as mottles/spicules, as well of
active region structures, such as arch filament systems (AFS),
filaments, fibrils, flaring regions, surges etc. Below, some
examples of such cloud model inversions are presented.

\subsection{Application to filaments}

Filaments are commonly observed features that appear on the solar
disc as dark long structures, lying along longitudinal magnetic
field inversion lines. When observed on the limb they are bright
and are called prominences. Filaments were some of the first solar
structures to be studied with cloud models \citep[see for
example][and references therein]{kt-malt76}. Since then several
authors used different cloud models to infer the dynamics and
physical parameters of filaments. \citet{kt-mein94}, for example,
studied the dynamical fine structure (threads) of a quiescent
filament assuming a number of identical -- except for the velocity
-- threads seen over the chromosphere and using a variant of BCM,
while \citet{kt-schm91} performed a similar study for threads by
using the DCM. \citet{kt-mori03} developed an interesting method
applying BCM to determine the three-dimensional velocity fields of
disappearing filaments.

\begin{figure}
  \centering
  \includegraphics[width=0.3\textwidth]{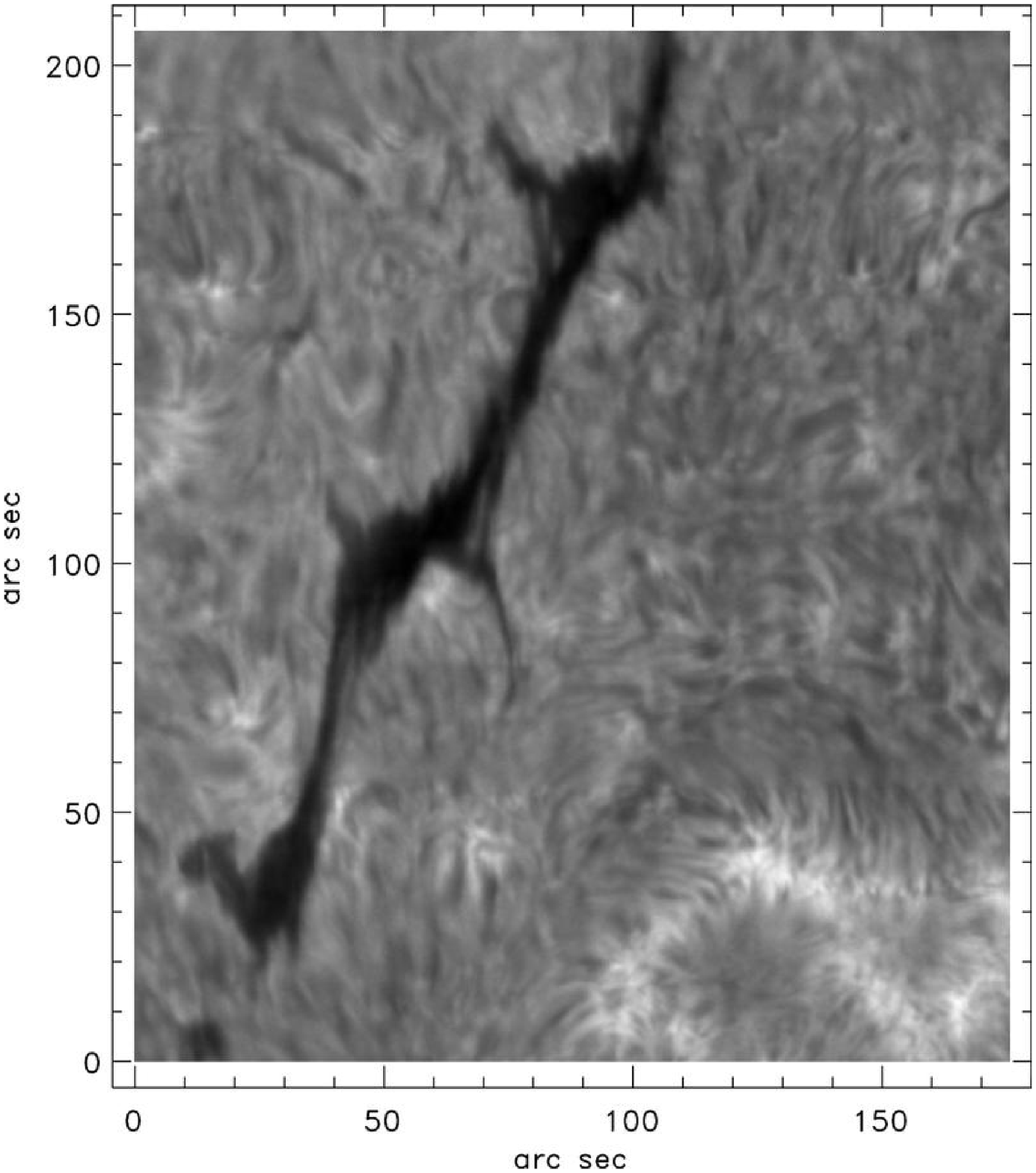}
  \includegraphics[width=0.3\textwidth]{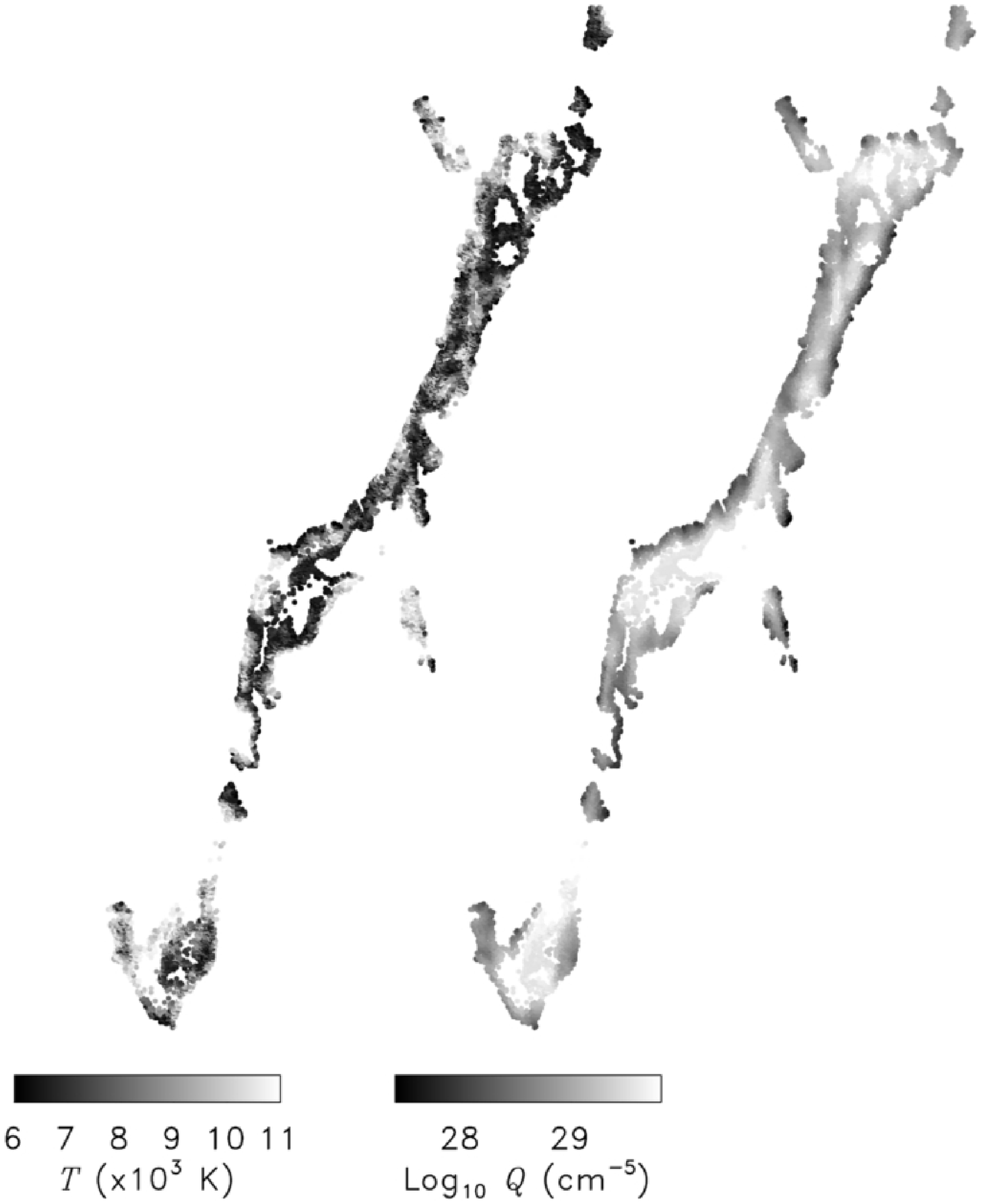}
  \includegraphics[width=0.3\textwidth]{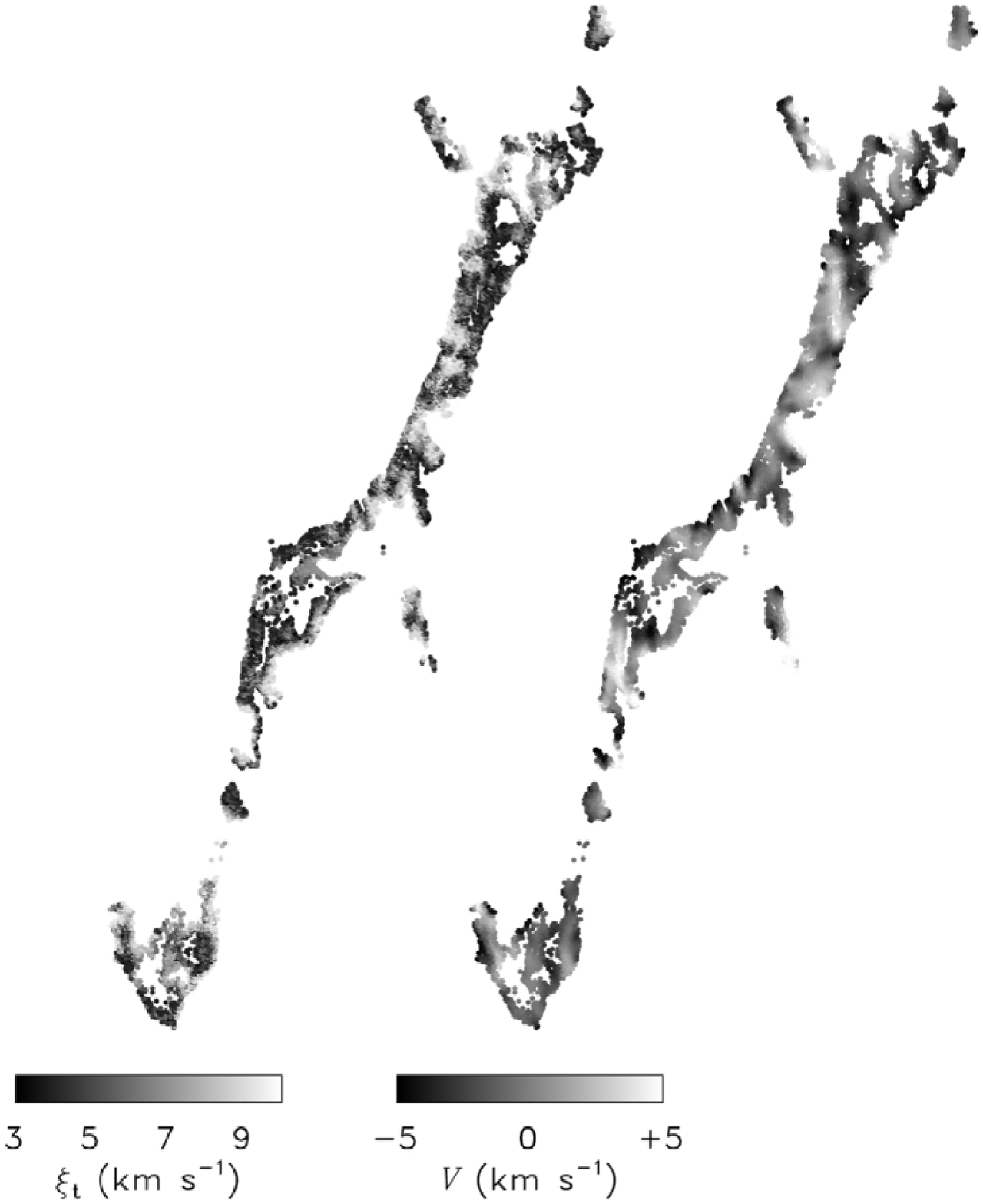}
  \includegraphics[width=0.3\textwidth]{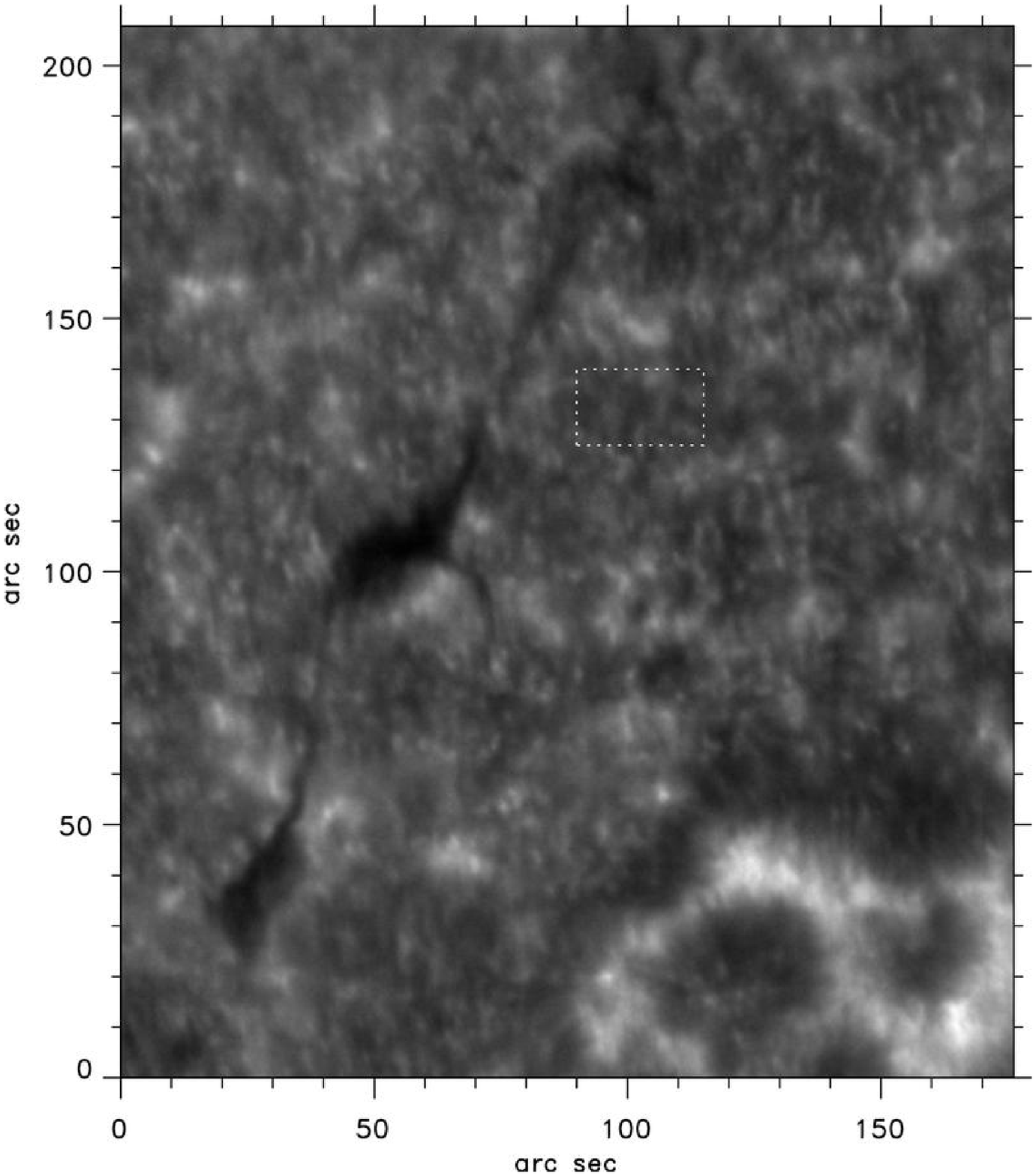}
  \includegraphics[width=0.33\textwidth]{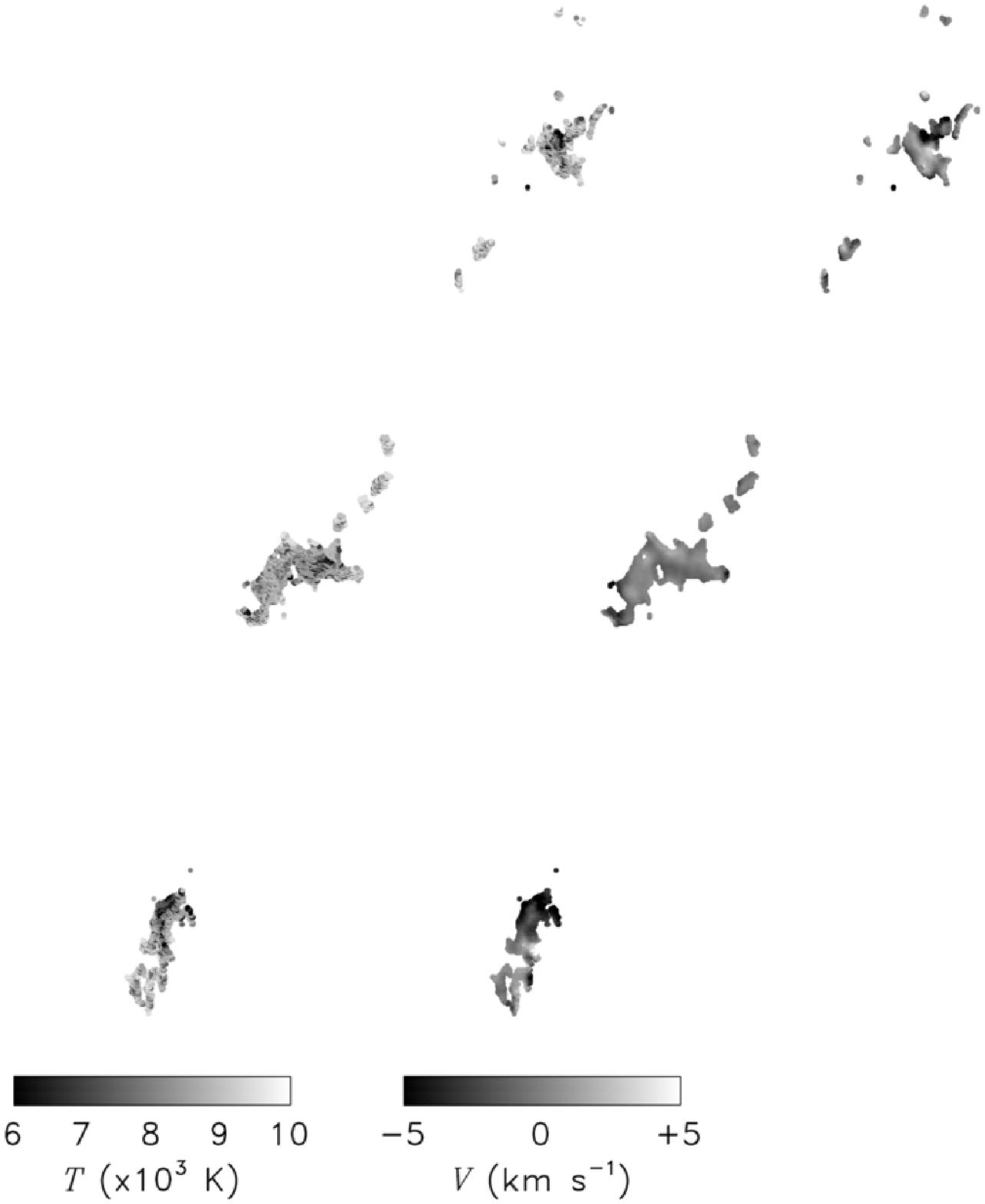}
  \includegraphics[width=0.33\textwidth]{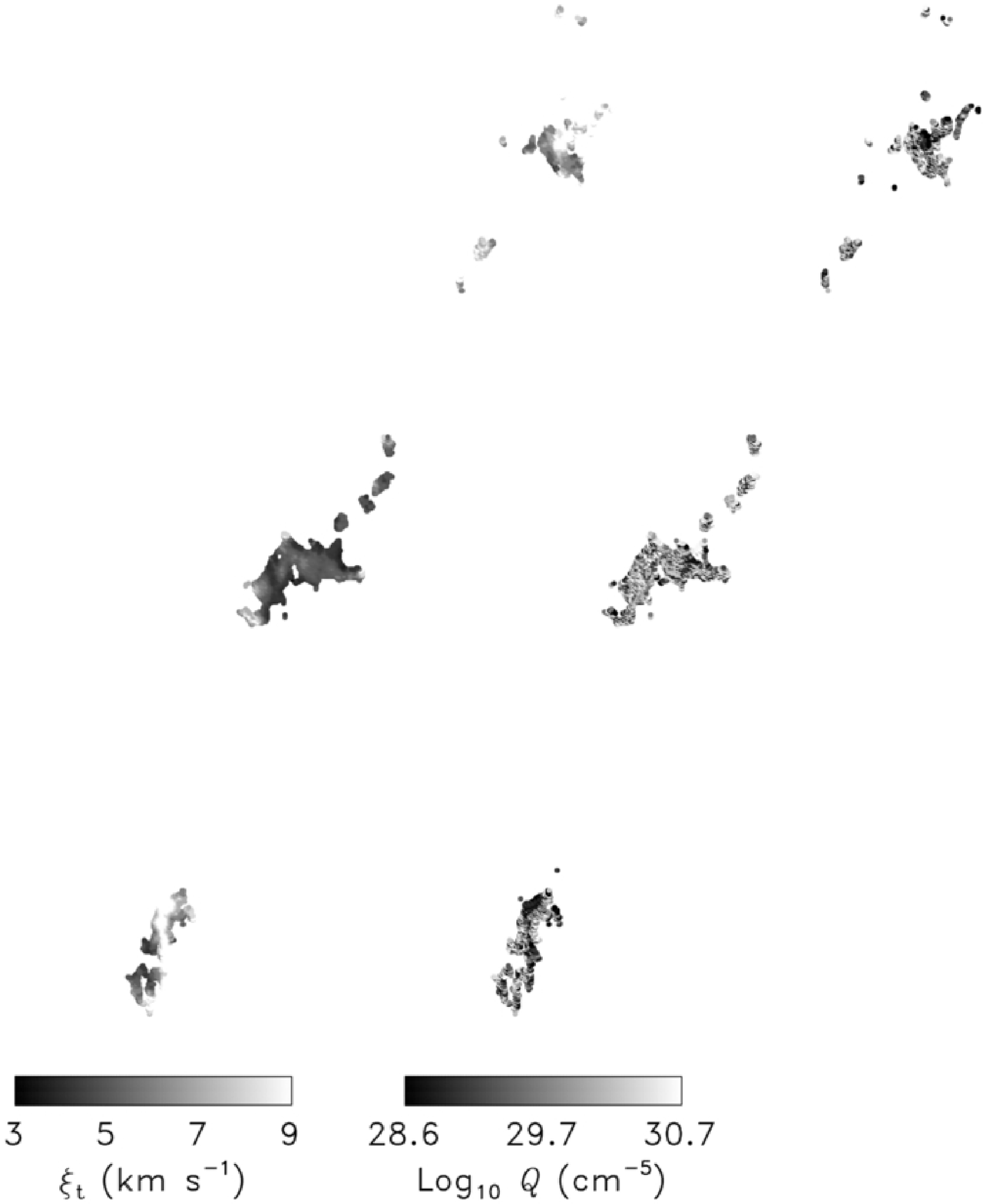}
  \caption[]{\label{kt-fig12}
  {\em Top row:} A filament observed in H$\alpha$ and the two-dimensional parameter distributions
  derived with a H$\alpha$  NLTE inversion using a grid of models. From \citet{kt-molo99}.
  {\em Bottom row:} Same filament observed in \ion{Ca}{ii} 8542\,\AA\, and the two-dimensional
   parameter distributions derived with a \ion{Ca}{ii} 8542\,\AA\, NLTE grid model inversion. From \citet{kt-tzio01}.
}\end{figure}

\citet{kt-molo99} and \citet{kt-tzio01} studied the same filament
observed in H$\alpha$ and \ion{Ca}{ii} 8542\,\AA\, respectively with the Multichannel
Subtractive Double Pass (MSDP) spectrograph
\citep{kt-mein91,kt-mein02} mounted on the German solar telescope
VTT in Tenerife. The filament was studied by using two very large
grids of models in H$\alpha$ and \ion{Ca}{ii} 8542\,\AA\, respectively which were constructed
with the NLTE one-dimensional code MALI \citep{kt-hein95}, as
described in Sect.~\ref{kt-nlte} Two-dimensional distributions of
the physical parameters were obtained (see Fig.~\ref{kt-fig12})
which are not that similar due to the different physical formation
properties and formation heights of the two lines.
\citet{kt-schm03} performed a similar NLTE grid inversion of a
filament combined with a classical BCM inversion in a
multi-wavelength study of filament channels. More recently,
\citet{kt-chae06} used H$\alpha$ images obtained with a tunable filter
and a BCM inversion to obtain detailed two-dimensional
distributions of the physical parameters describing a quiescent
filament.

\subsection{Application to arch filaments (AFS)}

\begin{figure}
  \centering
  \includegraphics[width=0.18\textwidth]{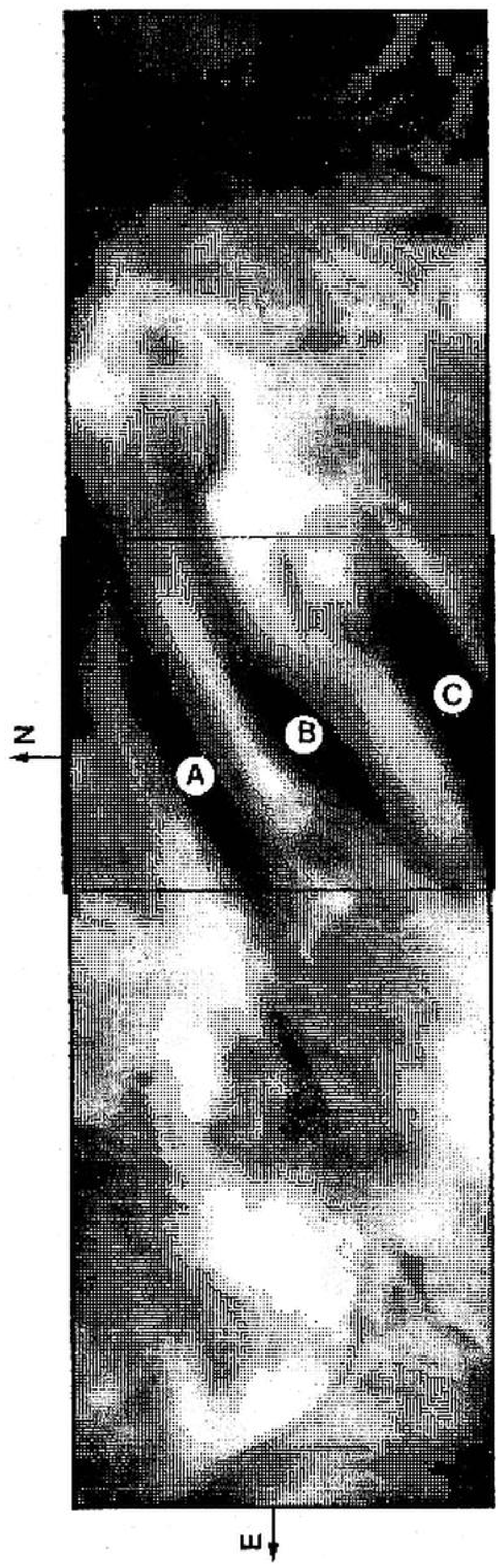}
  \includegraphics[width=0.4\textwidth]{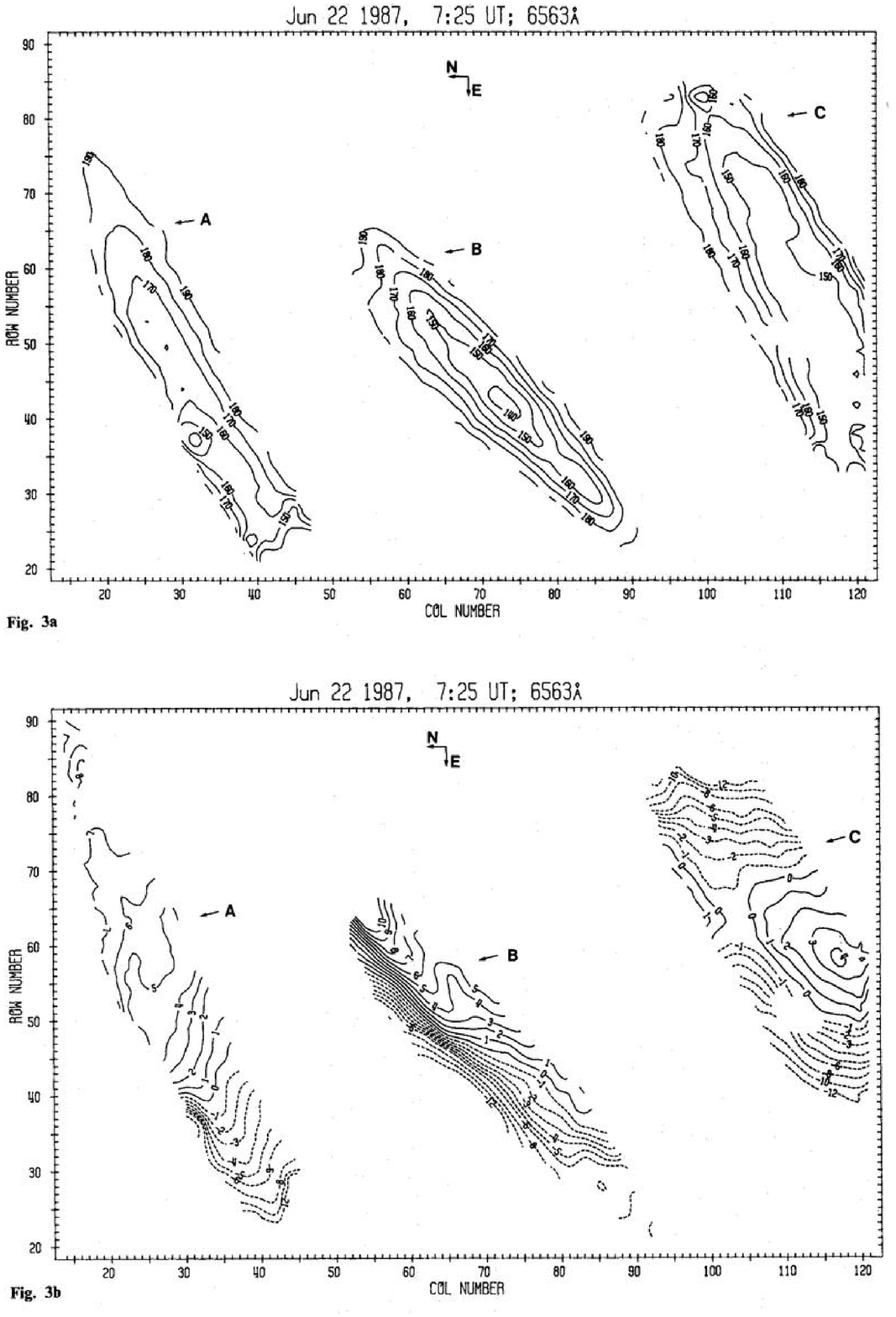}
  \caption[]{\label{kt-fig7}
  Contours maps of source function ({\em top right panel}) and the velocity ({\em bottom right panel}) derived
  with the cloud model for the AFS shown in H$\alpha$ in the left panel of the figure. From \citet{kt-alis90}.
}\end{figure}

Arch filaments systems (AFSs) are low-lying dark loop-like
structures formed during the emergence of solar magnetic flux in
active regions. \citet{kt-geor90} have used the Doppler signal
method described in Sect.~\ref{kt-dsmethod} to study mass motions
in AFSs observed in H$\alpha$, while \citet{kt-alis90} and
\citet{kt-tsir92} used the standard BCM to obtain the physical
parameters describing arch filament regions observed in the same
line (see Fig.~\ref{kt-fig7}). An example of the use of the
differential cloud model described in Sect.~\ref{kt-dcm} for the
study of the dynamics of AFSs can be found in \citet{kt-mein96b}
who applied the method to H$\alpha$ observations from a two-telescope
coordinated campaign. Finally \citet{kt-mein00b} present a study
of AFSs in \ion{Ca}{ii} 8542\,\AA\, using a fitting done with NLTE synthetic profile
calculations -- as described in Sect.~\ref{kt-nlte} -- with the
one-dimensional MALI code \citep{kt-hein95}.

\subsection{Application to fibrils}

\begin{figure}
  \centering
  \includegraphics[width=0.5\textwidth]{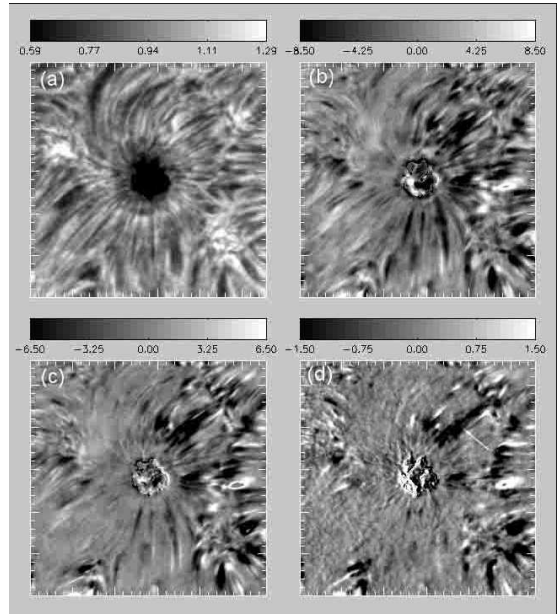}
  \caption[]{\label{kt-fig8}
  Image of a sunspot observed in H$\alpha$ (a) and Doppler velocity maps
  computed with the Doppler signal method from filtergrams in H$\alpha$$\pm$0.35\AA\ (b), in H$\alpha$$\pm$0.5\AA\ (c) and
  in H$\alpha$$\pm$0.75\AA\ (d). The intensity gray scale bar corresponds to normalized intensities while the
  Doppler velocity gray scale bars to velocities in~km~s$^{-1}$. From \citet{kt-geor03}.
}\end{figure}

Fibrils are small dark structures, belonging to the family of
``chromospheric fine structures", found in active regions
surrounding plages or sunspots (penumbral fibrils). One of the
first studies of fibrils was conducted by \citet{kt-bray74} who
compared observed profiles of fibrils with profiles calculated
with BCM. \citet{kt-alis90} used the standard BCM to obtain
two-dimensional maps of several physical parameters distributions
describing fibrils using H$\alpha$ observations obtained at Pic du Midi
Observatory. \citet{kt-geor03} used filtergrams obtained at nine
wavelengths along the H$\alpha$ to study the Evershed flow in sunspots
and reconstruct the three-dimensional velocity vector using the
Doppler signal method (see Fig.~\ref{kt-fig8}), while
\citet{kt-tsir00} used also the Doppler signal method to determine
LOS velocities of dark penumbral fibrils.

\subsection{Application to mottles}

Mottles are small-scale structures (appearing both bright and
dark) belonging also to the family of ``chromospheric fine
structures" and occurring at quiet Sun regions at the boundaries
of supergranular cells. Mottles are believed to be the
counterparts of limb spicules. They form groups called chains
(when they are almost parallel to each other) or rosettes (when
they are more or less circularly aligned, pointing radially
outwards from a central core) depending on their location at the
chromospheric network.

First cloud studies of mottles started with a controversy about
the ability of BCM to explain their contrast profiles.
\citet{kt-bray73} and \citet{kt-loug73} who studied bright and
dark mottles found that their contrast profiles are in good
agreement with BCM. However, \citet{kt-loug73} used also BCM to
deduce that it could not explain the contrast of individual bright
and dark mottles observed in H$\alpha$ near the limb, while
\citet{kt-cram75} claimed that the parameters inferred from an
application of BCM to contrast profiles of chromospheric fine
structures are unreliable.

\begin{figure}
  \centering
  \includegraphics[angle=1,width=0.45\textwidth]{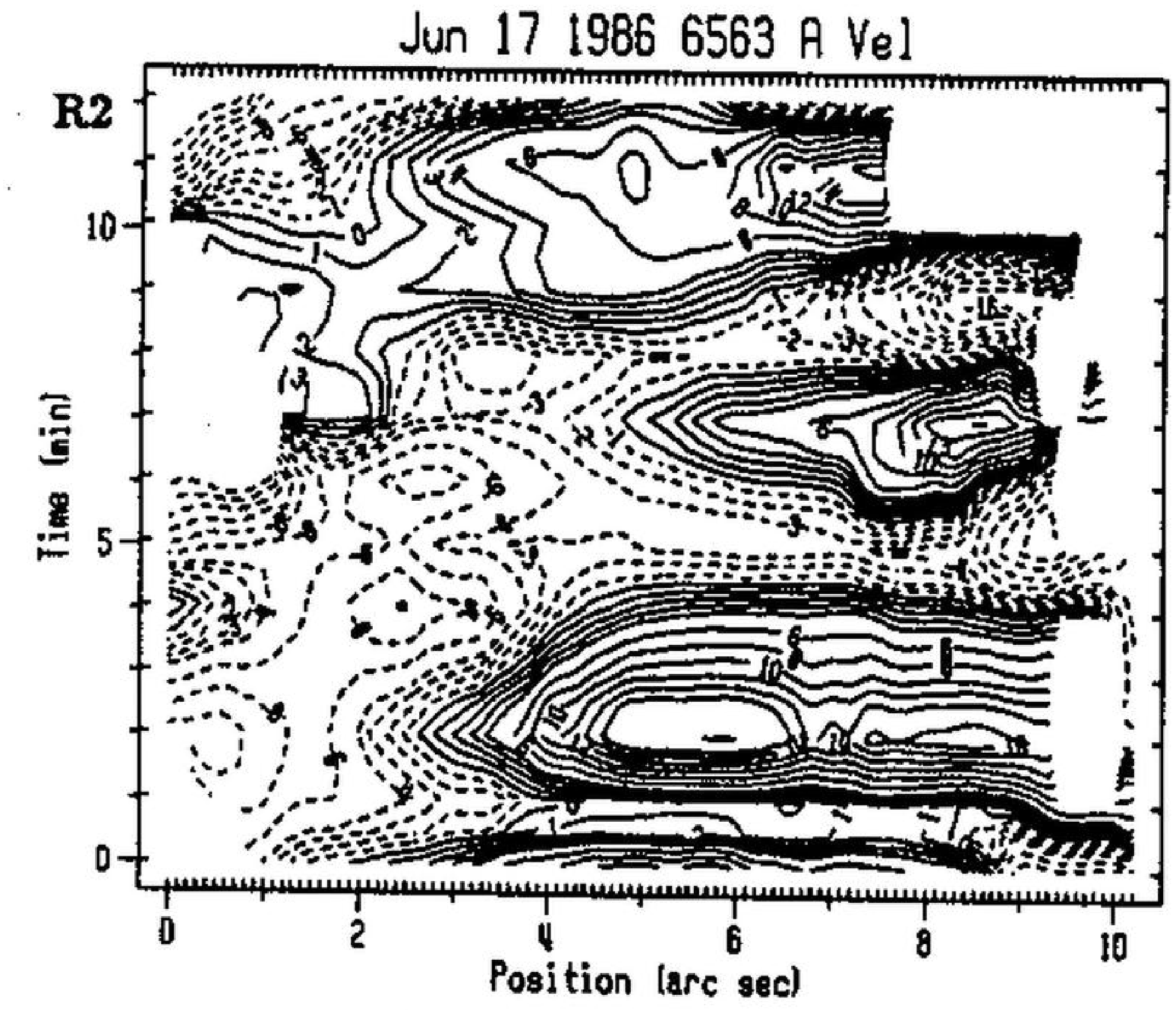}
  \includegraphics[width=0.45\textwidth]{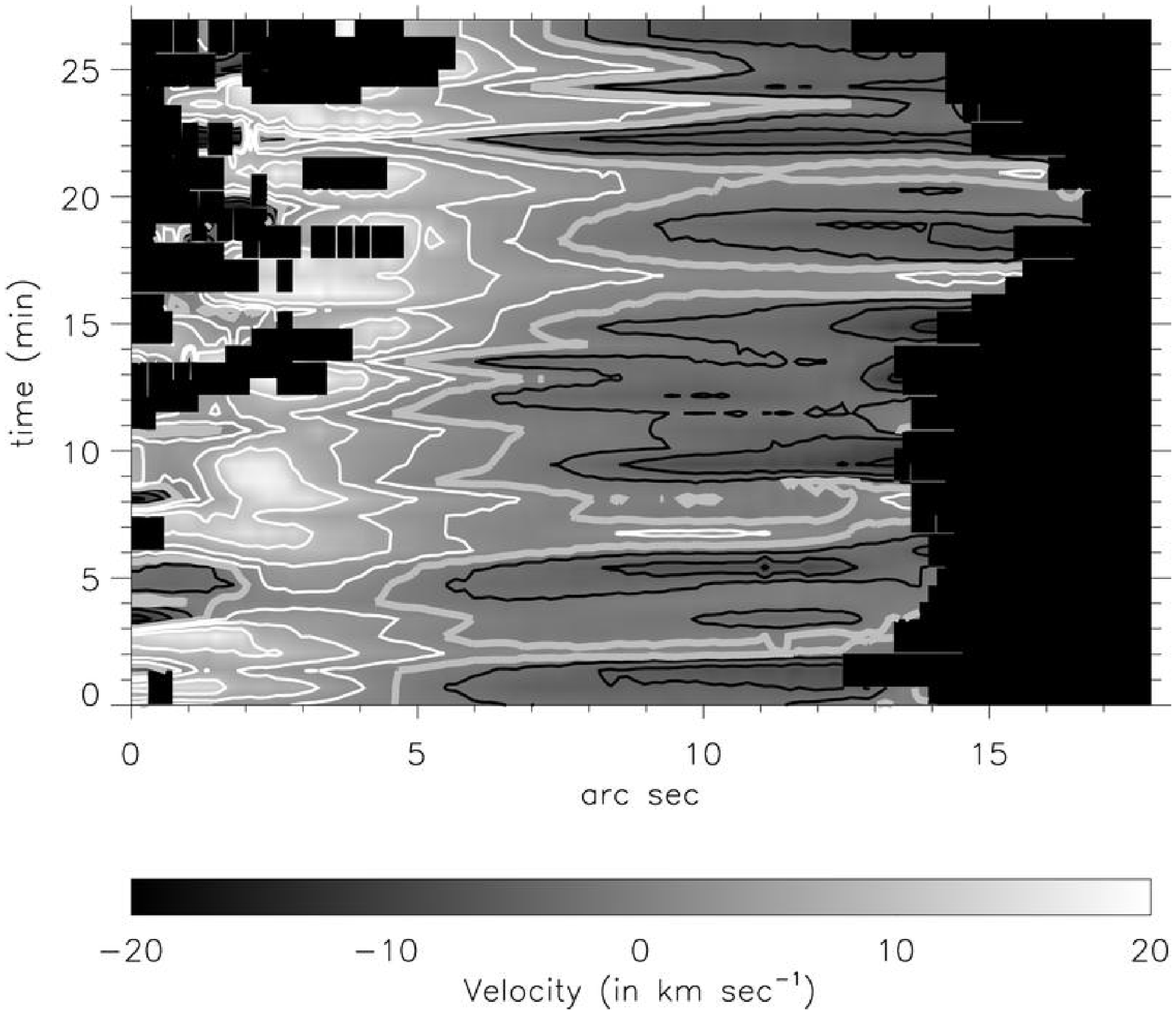}
  \caption[]{\label{kt-fig9}
  {\em Left panel:} Cloud velocity as a function of position and time
  along the axis of a dark mottle belonging to a rosette. From
  \citet{kt-tsir94}.  {\em Right panel:} Cloud velocity as a function
  of position and time along the axis of a dark mottle belonging to a
  chain of mottles. White contours denote downward velocities, black
  upward velocities, while the thick gray curve is the zero velocity
  contour. From \citet{kt-tzio03}.
}\end{figure}

Since then cloud models have been established as a reliable method
for the study of physical parameters of mottles. \citet{kt-tsir99}
studied several bright and dark mottles to derive physical
parameters assuming a constant as well as a varying source
function according to Eq.~\ref{kt-eq5}. \citet{kt-tsir97} applied
Beckers' cloud model to determine physical parameters in H$\alpha$ dark
mottles of a rosette region, while \citet{kt-tsir93,kt-tsir94}
studied the time evolution and fine structure of a rosette with
BCM and first showed an alternating behaviour with time for
velocity along mottles (see Fig.~\ref{kt-fig9}, {\em left panel}).
A similar behaviour has also been found by \citet{kt-tzio03} using
BCM for a chain of mottles (see Fig.~\ref{kt-fig9}, {\em right
panel}), while the dynamics of an enhanced network region were
also explored in high resolution H$\alpha$ images by \citet{kt-al04}.

\subsection{Application to post-flare loops}

\begin{figure}
  \centering
  \includegraphics[width=0.80\textwidth]{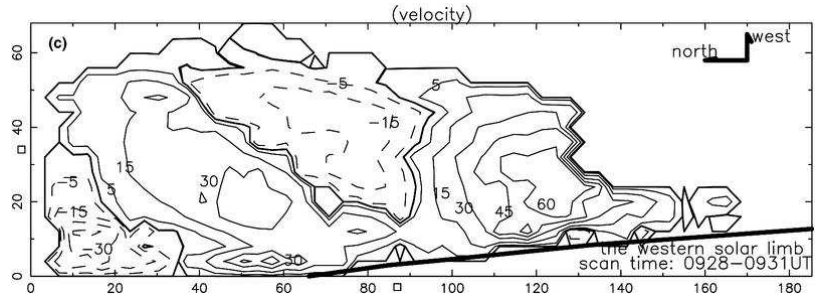}
  \caption[]{\label{kt-fig10}
  The distributions of Doppler velocity (in km~s$^{-1}$) derived with a
  multi-cloud method for H$\alpha$ limb post-flare loops. Coordinates
  are in units of arcsec, dashed curves show red-shifted mass motions,
  while solid curves indicate blue-shifted ones.  From
  \citet{kt-lian04}.
}\end{figure}

Post-flare loops are loops generally observed between two-ribbon
flares. We refer the reader to \citet{kt-bray83} for one of the
first post-flare loop studies, who constructed theoretical curves
based on the cloud model to fit observed contrast profiles of
active region loops. Later \citet{kt-schm88} and \citet{kt-hein92}
used a differential cloud model to study the structure and
dynamics of post-flare loops.  \citet{kt-hein92} also constructed
several isobaric and isothermal NLTE models of post-flare loops.
Their results were compared by \citet{kt-gu97} with
two-dimensional maps of H$\alpha$ post-flare loop cloud parameters
obtained using a two-cloud model. Multi-cloud models like the ones
described in Sect.~\ref{kt-mcmethod} were used by
\citet{kt-lian04} to study H$\alpha$ post-flare loops at the limb (see
Fig.~\ref{kt-fig10}), by \citet{kt-gu02} for the study of H$\alpha$ and
\ion{Ca}{ii} 8542\,\AA\, post-flare loops and by \citet{kt-dun00} for the study of {H$\beta$ }
post-flare loops. \citet{kt-liu01} obtained parameters of H$\alpha$
post-flare loops using the modified cloud model method presented
in Sect.~\ref{kt-nobg} that eliminates the use of the background
profile while \citet{kt-gu92} presented an extensive study using
BCM, the differential cloud model and a two-cloud model to study
the time evolution of post-flare loops in two-ribbon flares.
Finally, we refer the reader to \citet{kt-berl05} who studied H$\alpha$
ribbons during the gradual phase of a flare by comparing observed
H$\alpha$ profiles with a grid of synthetic H$\alpha$ profiles calculated with
the NLTE code MALI \citep{kt-hein95} which was modified to account
for flare conditions.

\subsection{Application to surges}

\begin{figure}
  \centering
  \includegraphics[width=0.2\textwidth]{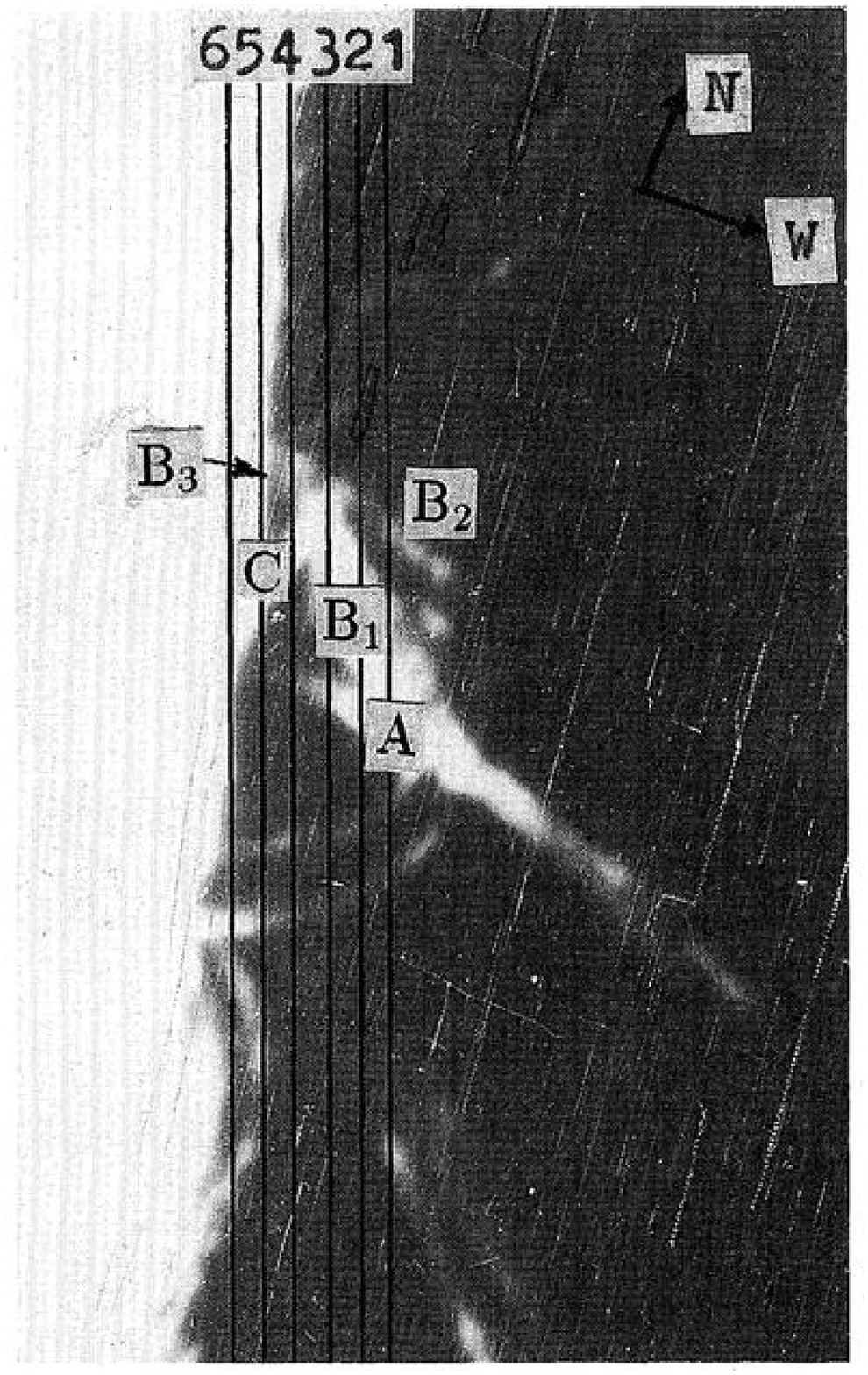}
  \includegraphics[width=0.45\textwidth]{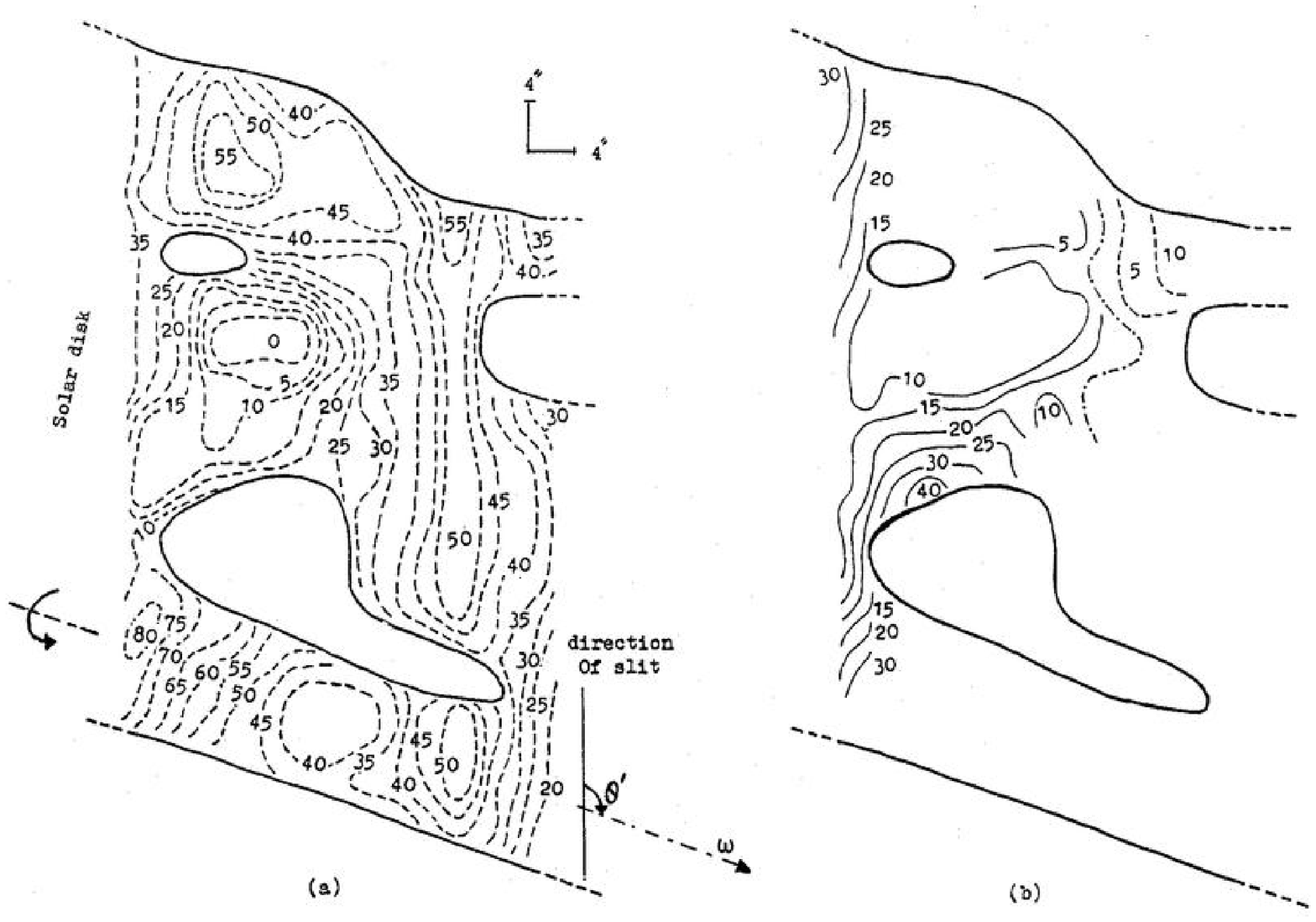}
  \caption[]{\label{kt-fig11}
  An H$\alpha$ filtergram of a surge ({\em left panle}) and the
  two-dimensional isocontours of Doppler velocity derived with a
  two-cloud model. Dashed curves refer to blue-shifted velocities
  ({\em middle panel}), solid curves red-shifted ones ({\em right panel}),
  while the unit of velocity is in km~s$^{-1}$. From \citet{kt-gu94}.
}\end{figure}

Surges are large jet-like structures observed in opposite polarity
flux emergence areas in active regions believed to be supported by
magnetic reconnection. \citet{kt-gu94} studied a surge on the limb
observed in H$\alpha$, using a two-cloud model inversion as described
in Sect.~\ref{kt-mcmethod} (see Fig~\ref{kt-fig11}). The inversion
result was detailed two-dimensional maps of the blue-shifted and
red-shifted LOS velocity distributions.

\section{Conclusions}

Several inversion techniques for chromospheric structures based on
the cloud model have been reviewed. Cloud models are fast, quite
reliable tools for inferring the physical parameters describing
cloud-like chromospheric structures located above the solar
photosphere and being illuminated by a background radiation. Cloud
model techniques usually provide unique solutions and the results
do not differ -- in principle -- qualitatively, especially for
velocity, when using different cloud model techniques. However
there can be quantitative differences arising from a) the
selection of the background intensity, b) the physical conditions
and especially the behaviour of the source function within the
structure under study, and c) the particular model assumptions.
Cloud models are mainly used for absorbing structures, however
most of the techniques do work also for line-center contrasts that
are slightly higher than zero, indicating an important emission by
the structure itself.

Several different variants for cloud modeling have been proposed
in literature so far that mainly deal with different assumptions
or calculations for the source functions and span from the simple
BCM that assumes a constant source function to more complicated
NLTE calculations of the radiation transfer and hence the source
function within the structure. Accordingly, several different
techniques -- most of them iterative -- have been proposed for
solving cloud model equations. The latest and more accurate
inversion techniques involve the construction of large grids of
synthetic profiles, for different geometries and physical
conditions, which are used for comparison with observed profiles.

Cloud models can be applied with success to several, different in
geometry and physical conditions, solar structures both of the
quiet Sun, as well as of active regions. The resulting parameter
inversions has shed light to several problems involving the
physics and dynamics of chromospheric structures.

The future of cloud modeling looks even more brighter. New high
resolution data from telescopes combined with an always increasing
computer power and the continuous development of new, state of the
art, NLTE one-dimensional and two-dimensional cloud model codes
will provide further detailed insights to the physics and dynamics
that govern chromospheric structures.\\

\acknowledgements KT thanks G.~Tsiropoula for constructive
comments on the manuscript and acknowledges support by the
organizers of the meeting and by Marie Curie European
Reintegration Grant MERG-CT-2004-021626.


\end{document}